\documentclass[cernpreprint, atlasdraft=false, texlive=2023, UKenglish, texmf, orcidlogo]{atlasdoc}
\usepackage{atlaspackage}
\usepackage{atlasbiblatex}

\usepackage{atlasphysics}

\usepackage{bm}

\usepackage{enumerate}
\usepackage{multirow}
\usepackage{tablefootnote}

\graphicspath{{logos/}{figures/}}

\usepackage{ANA-EXOT-2018-64-PAPER-defs}

\AtlasTitle{Combination and summary of ATLAS dark matter searches interpreted in a 2HDM with a pseudo-scalar mediator using 139~fb$^{-1}$~of $\sqrt{s} = 13$~TeV $pp$ collision data}
\AtlasAbstract{%
Results from a wide range of searches targeting different experimental signatures with and without missing transverse momentum ($E_{\mathrm{T}}^{\mathrm{miss}}$)
are used to constrain a Two-Higgs-Doublet Model (2HDM) with an additional pseudo-scalar mediating the interaction between ordinary and dark matter (2HDM+$a$).
The analyses use up to 139~fb$^{-1}$ of proton--proton collision data at a centre-of-mass energy $\sqrt{s}=13$~TeV recorded with the ATLAS detector
at the Large Hadron Collider during 2015--2018.
The results from three of the most sensitive searches are combined statistically.
These searches target signatures with large $E_{\mathrm{T}}^{\mathrm{miss}}$ and a leptonically decaying $Z$ boson;
large $E_{\mathrm{T}}^{\mathrm{miss}}$ and a Higgs boson decaying to bottom quarks; and production of charged Higgs bosons in final states with top and bottom quarks, respectively.
Constraints are derived for several common and new benchmark scenarios in the 2HDM+$a$.
}

\AtlasRefCode{EXOT-2018-64}

\PreprintIdNumber{CERN-EP-2023-088}

\AtlasJournal{Science Bulletin}
\AtlasJournalRef{Science Bulletin 69 (2024) 3005}
\AtlasDOI{https://doi.org/10.1016/j.scib.2024.06.003}

\hypersetup{pdftitle={ATLAS document},pdfauthor={The ATLAS Collaboration}}

\begin{document}

\maketitle

\tableofcontents

\section{Introduction}
%
%

% The next lines are included from the .//sections/introduction.tex input file
The existence of dark matter (DM) is supported by a plethora of astrophysical measurements,
including the rotational speed of stars in galaxies~\cite{Corbelli:1999af,Rubin:1980,Begemann:1991},
precision measurements of the cosmic microwave background~\cite{Hinshaw:2012aka,Akrami:2018vks},
and gravitational lensing measurements~\cite{Trimble1987,Bertone2005,Feng2010}.
However, little is known of its particle nature, which remains one of the central questions in particle physics.
The particle content of the Standard Model (SM) is insufficient to explain these observations;
thus, a satisfactory dark matter candidate is a strong consideration in many Beyond-the-SM (BSM) extensions.

Complementary probes of DM are underway in several areas, from indirect searches for the products of dark matter annihilation or decay~\cite{Acharyya:2023ptu,HESS:2022ygk,MAGIC:2021mog,CTA:2020qlo,HAWC:2017udy,Aartsen:2018mxl,IceCube:2021xzo,Fermi-LAT:2016uux,TheFermi-LAT:2017vmf},
searches for the direct detection of DM scattering elastically off nuclei and electrons~\cite{Aalbers:2022dzr,LZ:2022ufs,Akerib:2016vxi,PICO:2019vsc,PandaX-II:2021nsg,Cui:2017nnn,XENON:2022ltv,XENON:2023sxq,PhysRevLett.122.141301,PhysRevLett.123.241803,DarkSide-50:2022qzh,Angloher:2015ewa,Lai:2023qub,Amaudruz:2017ekt,Agnese:2017njq,Agnese:2017jvy},
and recent searches using gravitational-wave interferometers~\cite{LIGOScientific:2021ffg,Vermeulen:2021epa},
to searches for the production of dark matter at collider experiments, such as the ATLAS experiment~\cite{PERF-2007-01} at the Large Hadron Collider (LHC)~\cite{Evans:2008zzb}.
General purpose particle physics experiments are sensitive to a wide variety of potential dark matter candidates,
such as axions~\cite{Georgi:165431,Choi:1986zw,Preskill:1982cy,Abbott:1982af,Dine:1982ah} or Weakly Interacting Massive Particles (WIMPs)~\cite{Steigman:1984ac}.
The motivation for the latter arises from a paradigm known as the WIMP miracle~\cite{Steigman:1984ac}
Assuming DM to be produced via the freeze-out mechanism, the relic density of non-relativistic matter in the early universe~\cite{Kolb:1990vq},
measured in data from the WMAP~\cite{Hinshaw:2012aka} and Planck~\cite{Akrami:2018vks} missions,
can be achieved when the DM mass is close to the electroweak scale and when the DM coupling to Standard Model particles is of the order of the weak interaction.
Consequently, WIMP DM particles could be produced and studied at the LHC experiments.

A particular strength of collider searches lies in the fact that the high-energy collisions of SM particles could not only produce DM directly under controlled experimental conditions
but also provide access to particles mediating the interactions between DM and the SM sector.
A \textit{mediator} produced in a collision could decay into DM particles, which themselves could not be detected,
resulting in a momentum imbalance in the plane tranverse to the collision axis, referred to as missing transverse momentum \ptmiss, with magnitude \met.
Alternatively, a mediator could decay back into SM particles, from which its properties could be reconstructed.

Dark matter searches at the LHC explore both these avenues in the quest to solve the puzzle of DM.
Invisible mediator decays can be detected only if the mediator is produced in association with another particle or particles,
for example a quark or gluon from initial-state radiation that results in a hadronic jet ($j$), leading to a characteristic $\met+j$ signature~\cite{EXOT-2018-06,CMS-EXO-20-004}.
These signatures are referred to as $\met+X$ signatures in the following.
Visible mediator decays allow for the reconstruction of the mediator particle from its decay products, for example in the context of resonance searches,
if the mediator is produced in the $s$-channel~\cite{EXOT-2019-03,EXOT-2018-08,CMS-EXO-16-056,CMS-EXO-19-012,CMS-EXO-19-019}.

The searches mentioned above are traditionally interpreted in the context of \textit{simplified models} of DM,
which rely on a minimal set of new particles and interactions.
The most commonly used among these simplified models postulate the existence of a single fermionic DM particle and a single mediator, which, depending on the model,
may be a vector, axial-vector, scalar, or pseudo-scalar particle~\cite{Abercrombie:2015wmb,Boveia:2016mrp,Albert:2017onk}.
The models are characterised by a minimal set of free parameters, typically the masses and couplings of the DM and mediator particles.
While this facilitates the definition of benchmark scenarios that can be used to compare results between experiments,
the theoretical incompleteness of simplified models can limit the range of collider signatures realised.

A more complete benchmark model with a rich collider phenomenology is explored in this paper,
known as the Two Higgs Doublet Model (2HDM) plus pseudo-scalar mediator $a$, denoted 2HDM+$a$~\cite{Bauer:2017ota}.
In this model, the scalar sector of the SM is extended by an additional complex doublet,
an extension that is well motivated by theories beyond the SM addressing, for example,
the electroweak hierarchy problem~\cite{Golfand:1971iw,Volkov:1973ix,Wess:1974tw,Wess:1974jb,Ferrara:1974pu,Salam:1974ig},
baryogenesis~\cite{Dorsch:2016nrg,McLerran:1990zh,Turok:1990zg,Cohen:1991iu,Cline:1996mga,Fromme:2006cm,Cline:2011mm},
or the strong CP problem~\cite{Kim:1986ax}.
The model also contains a pseudo-scalar mediator which couples to a fermionic dark matter candidate, $\chi$.

The 2HDM+$a$ is a simple, ultra-violet-complete (UV-complete), gauge-invariant, and renormalisable extension of the
pseudo-scalar mediator simplified models~\cite{Buckley:2014fba,Abercrombie:2015wmb}.
A pseudo-scalar mediator is chosen primarily due to the reduced constraints from direct detection experiments,
and its ability to reproduce the observed relic abundance across much of the model parameter space, making LHC searches particularly important.
Another reason the \thdma is of high interest for the LHC community is the fact that it predicts a wide range of collider signatures with a complex interplay across the model parameter space,
including signatures not predicted in the commonly used simplified models.
It is promoted by the LHC Dark Matter Working Group as a complete benchmark model~\cite{2HDMWGproxi}.

In total, the \thdma adds five new states to the SM scalar sector: a scalar $H$, pseudo-scalar $A$, charged Higgs bosons H$^{\pm}$, and the pseudo-scalar mediator $a$.
After the discovery of the Higgs boson $h$ by the LHC experiments~\cite{HIGG-2012-27,CMS-HIG-12-028}, the exploration of the scalar sector of the SM is another high experimental priority.
The results of searches for additional (pseudo-)scalar bosons constrain this model, thus complementing constraints from searches targeting $\met+X$ signatures.

A comprehensive synopsis is presented of the diverse set of collider signatures of the \thdma benchmark explored through improved ATLAS searches using 139~fb$^{-1}$ of LHC Run 2 data.
Compared with earlier summaries, additional signatures are considered, the individual analysis exclusions are improved and a wider range of interpretations are considered.
In particular, a statistical combination is performed of three of the most sensitive analyses:
a search for large \met produced in association with a leptonically decaying $Z$ boson, \monozll~\cite{HIGG-2018-26},
a search for large \met produced in association with a SM Higgs boson decaying into $b\bar{b}$, \monohbb~\cite{EXOT-2018-46},
and a search for associated production of a top and a bottom quark with a charged Higgs boson decaying into a top and a bottom quark, \htb~\cite{HDBS-2018-51}.
For the first time, constraints from searches targeting the \monojet~\cite{EXOT-2018-06}, $\met+tW$~\cite{EXOT-2021-01}, \monohtautau~\cite{HDBS-2018-50},
and \hlrs~\cite{HDBS-2021-03,HIGG-2017-05,HIGG-2014-02,EXOT-2016-22,HDBS-2018-55} signatures are included in the summary,
besides constraints from searches targeting the \monohgamgam~\cite{HIGG-2019-02}, \tttt~\cite{EXOT-2019-26}, and \hinv~\cite{HIGG-2021-05} signatures.

In addition to the usual gluon--gluon ($gg$) initiated processes, $b\bar{b}$-initiated production is considered for all relevant signatures,
which is dominant in some regions of the model parameter space.
A full set of the benchmark scenarios recommended in Ref.~\cite{2HDMWGproxi} is featured,
with an updated definition for the interpretation varying the DM mass motivated by the increased sensitivity of the searches.
Finally, a new scenario is introduced, following Ref.~\cite{Argyropoulos:2022ezr}, to showcase possibilities for lighter pseudo-scalar mediators,
and the interplay of light resonance searches with the \met\ signatures.

A previous ATLAS summary paper included constraints on the \thdma benchmark from dark matter searches using 36~fb$^{-1}$ of $\sqrt{s}=13$~\TeV proton--proton ($pp$)
collision data~\cite{EXOT-2017-32}.
Constraints on the model have also been placed by the CMS Collaboration using searches in the \monohbb~\cite{CMS-EXO-18-011}
and \monozll~\cite{CMS-EXO-19-003} final states with 137~fb$^{-1}$ of LHC Run 2 data.

The paper is organised as follows: In Section~\ref{sec:thdma}, the theoretical set-up, benchmark model and choice of scenarios are discussed in detail;
in Section~\ref{sec:detector} the ATLAS detector is described;
details of the signal and background modelling are given in Section~\ref{sec:samples}.
In Section~\ref{sec:analyses} brief overviews of the experimental signatures and the analyses targeting them are described;
details of systematic uncertainties and the statistical combination of analyses are given in Sections~\ref{sec:systematics} and \ref{sec:comb};
the combined results and summaries of the experimental constraints can be found in Section~\ref{sec:summary};
a summary of the findings is given in Section~\ref{sec:conclusion}.

% End of text imported from the .//sections/introduction.tex input file

%
\section{Theoretical framework}
\label{sec:thdma}
%

% The next lines are included from the .//sections/thdma.tex input file
The benchmark model used in this publication builds on the assumption of the existence of a second complex Higgs doublet,
which is postulated in various UV-complete theories with an extended Higgs sector.
The 2HDM sector is assumed to have a CP-conserving potential and a softly broken $\mathbb{Z}_2$ symmetry~\cite{Gunion:2002zf}.
After electroweak symmetry breaking, the 2HDM contains five Higgs bosons:
a lighter CP-even boson, $h$, a heavier CP-even boson, $H$, a CP-odd boson, $A$, and two charged bosons, $H^{\pm}$.
The 2HDM coupling structure is chosen to be of type-II~\cite{Gunion:2002zf} and the alignment and decoupling limits are assumed,
so that the lighter CP-even state $h$ can be identified with the SM Higgs boson.
In addition, the model includes a fermionic DM particle $\chi$ and a pseudo-scalar (CP-odd) mediator $a$ with Yukawa-like couplings
to both the SM fermions and the Dirac DM particle $\chi$, thus allowing for interactions between DM and the SM sector.
The mediator mixes with the pseudo-scalar $A$ of the 2HDM sector with mixing angle $\theta$.

\begin{figure}[h!]
\centering
\subfloat[]{\includegraphics[width=0.26\textwidth]{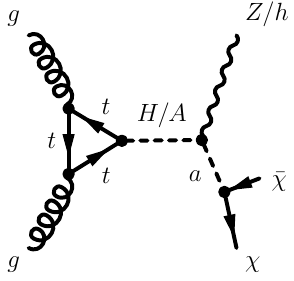}
\label{fig:THDMpa_feyn_gg_res}
}
\subfloat[]{\includegraphics[width=0.28\textwidth]{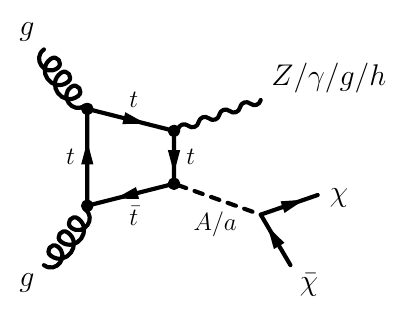}
\label{fig:THDMpa_feyn_gg_box}
}
\subfloat[]{\includegraphics[width=0.26\textwidth]{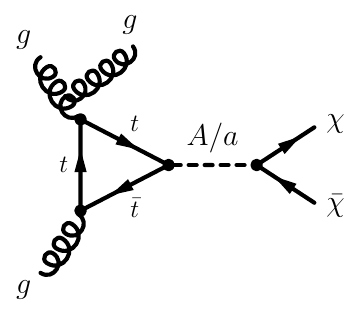}
\label{fig:THDMpa_feyn_gg_jet}
} \\
\subfloat[]{\includegraphics[width=0.28\textwidth]{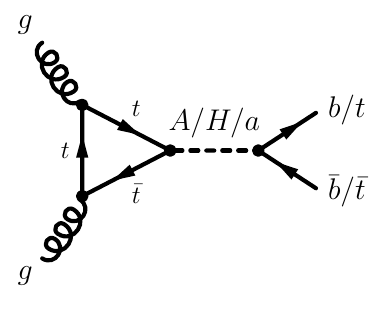}
\label{fig:THDMpa_feyn_tt_res}
}
\subfloat[]{\includegraphics[width=0.28\textwidth]{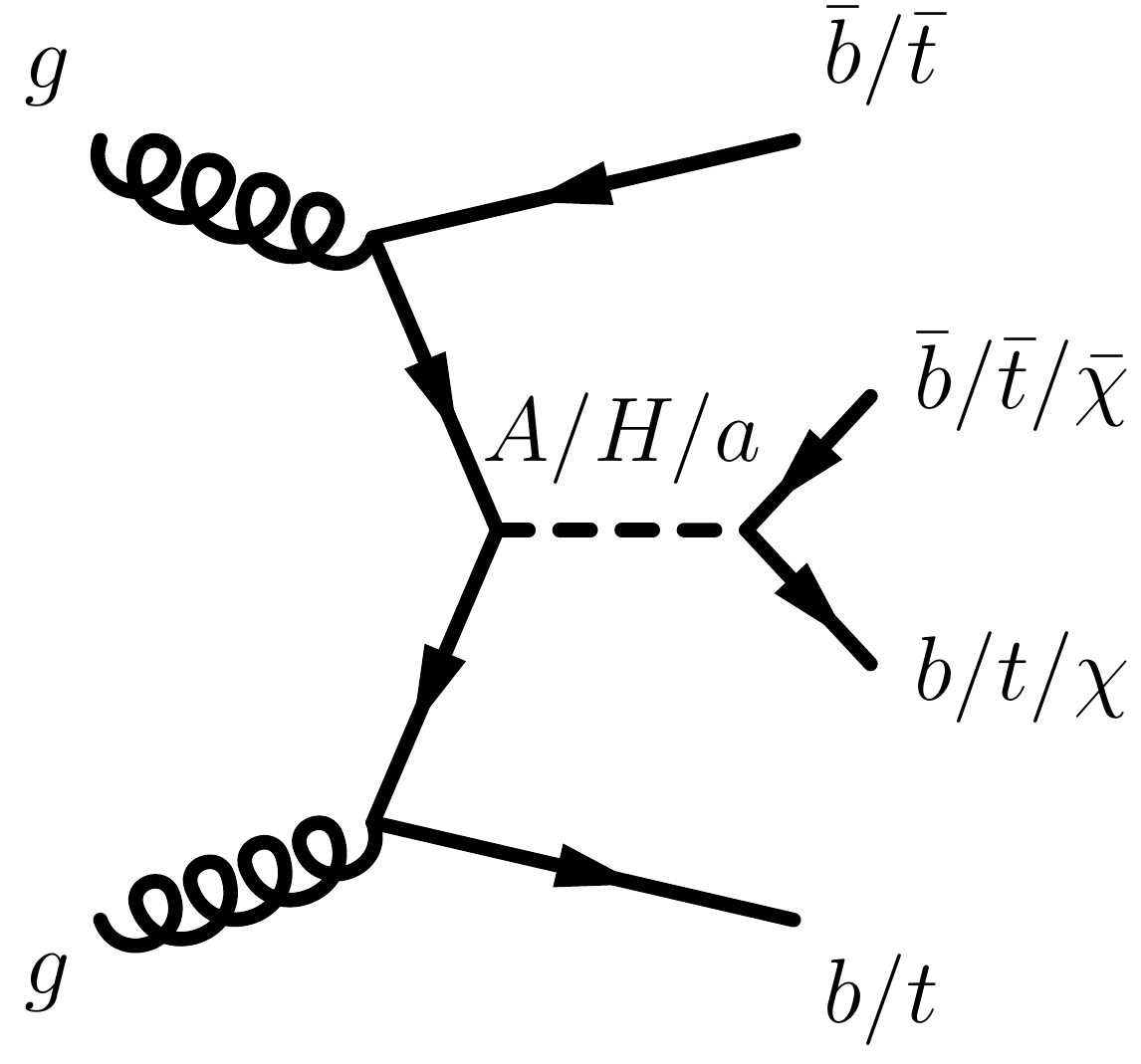}
\label{fig:THDMpa_feyn_res_assoc}
}\\
\subfloat[]{\includegraphics[width=0.22\textwidth]{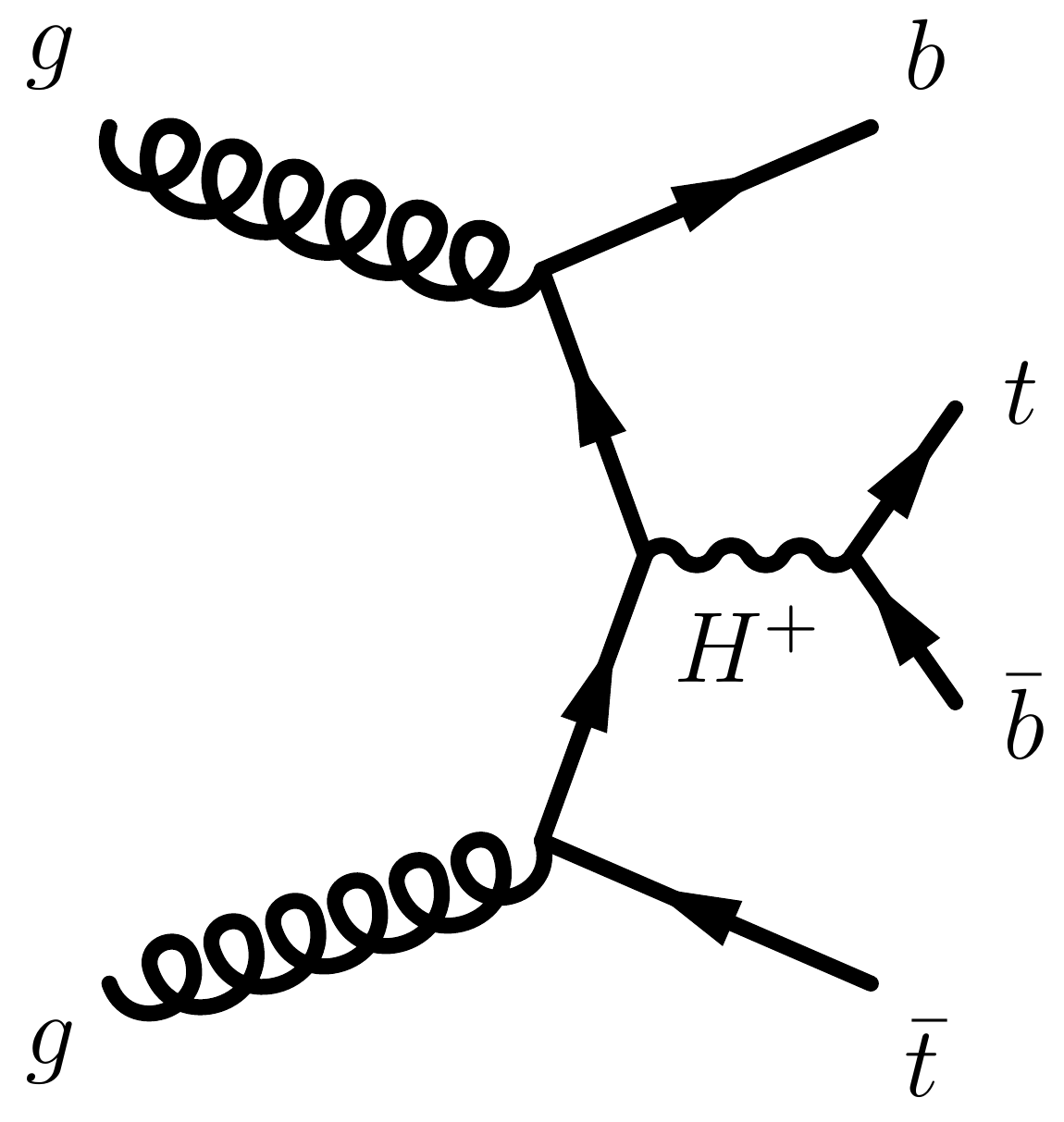}
\label{fig:THDMpa_feyn_htb}
}
\subfloat[]{\includegraphics[width=0.30\textwidth]{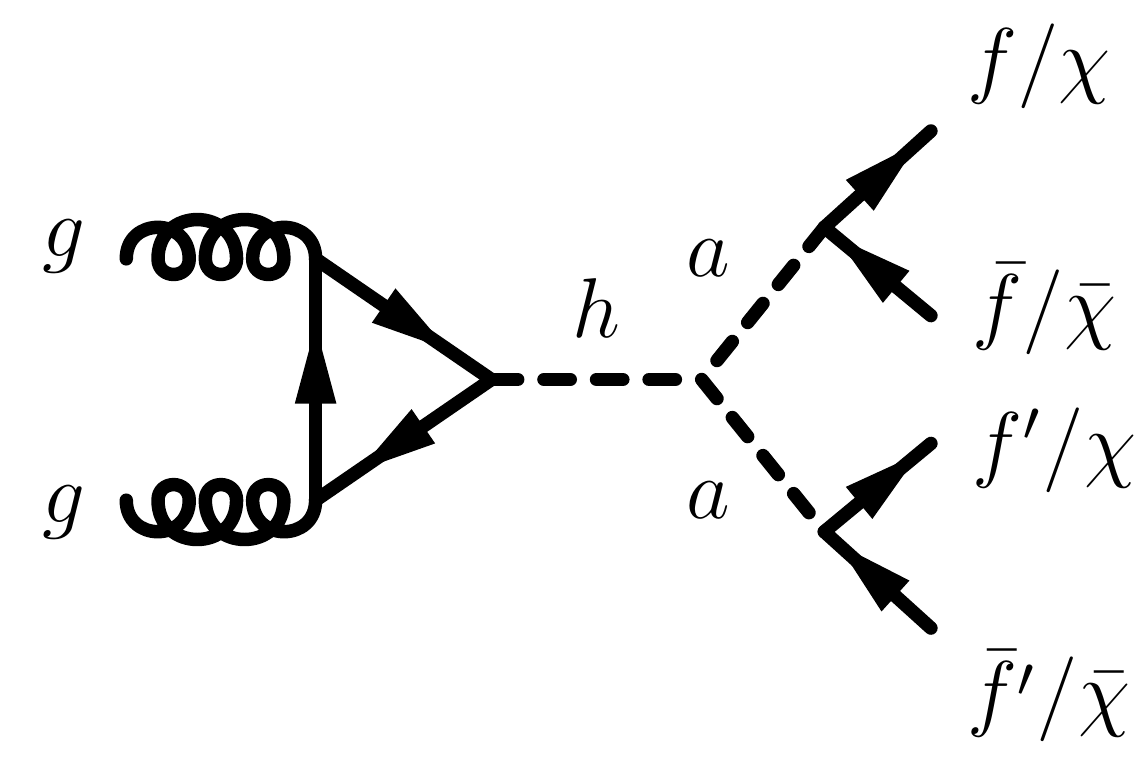}
\label{fig:THDMpa_feyn_haa}
}
\caption{
\label{fig:THDMpa_feyn_gg}
Representative Feynman diagrams for the dominant gluon-induced production and decay modes in the 2HDM+$a$:
(a,b) resonant and non-resonant production of the $\met+Z$ and $\met+h$ signatures,
(c) production of the $\met+j$ signature,
(d) resonant $A/H$ production with decay into \ttbar or \bbar,
(e) \ttbar or \bbar associated resonant $A/H$ production, leading to \tttt, $b\bar{b}b\bar{b}$, $t\bar{t}b\bar{b}$, $\met+\ttbar$, or $\met+\bbar$ signatures,
(f) \htb production,
and (g) production of a SM Higgs boson decaying into a pair of mediators $aa$ with subsequent decays into fermions or DM.}
\end{figure}

\begin{figure}[h!]
\centering
\subfloat[]{\includegraphics[width=0.35\textwidth]{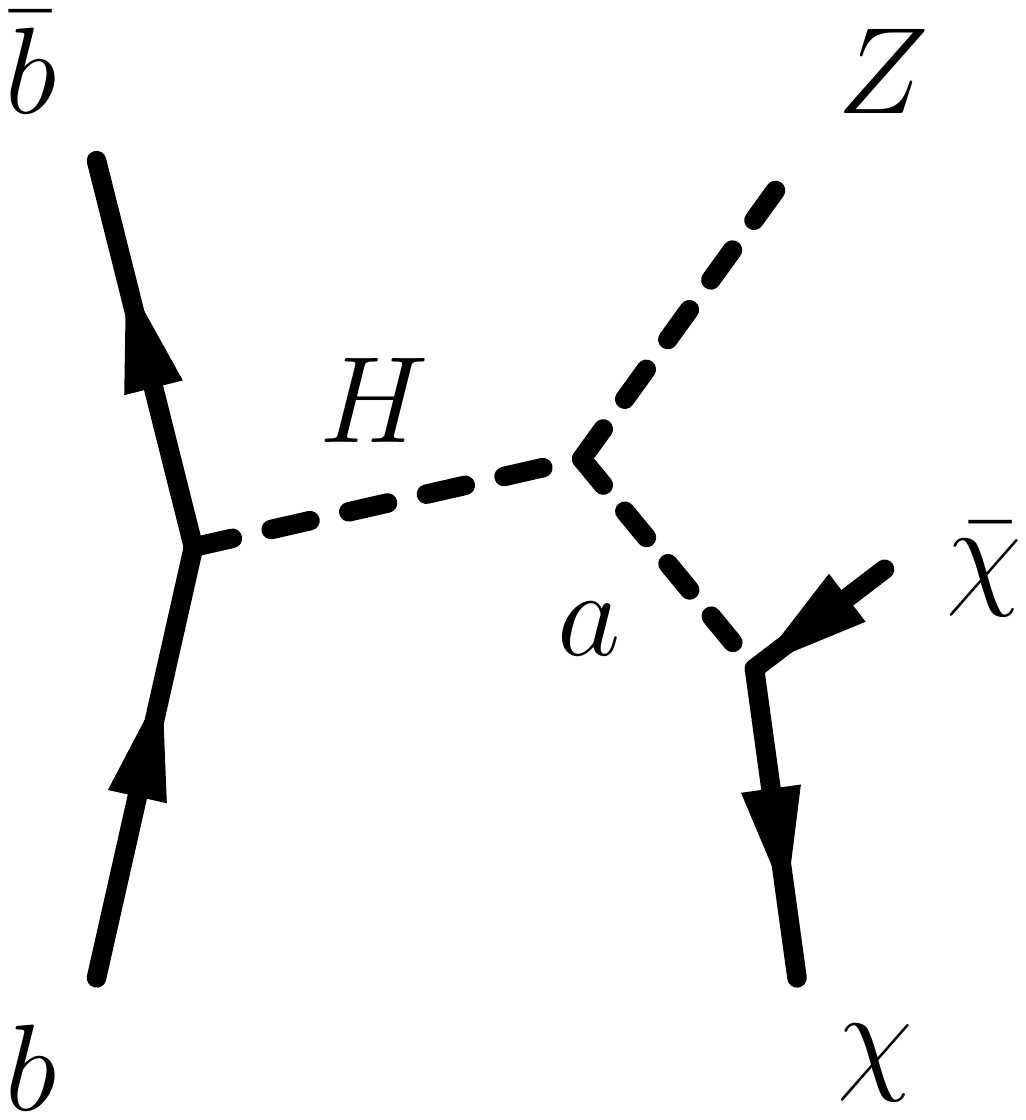}
}
\subfloat[]{\includegraphics[width=0.20\textwidth]{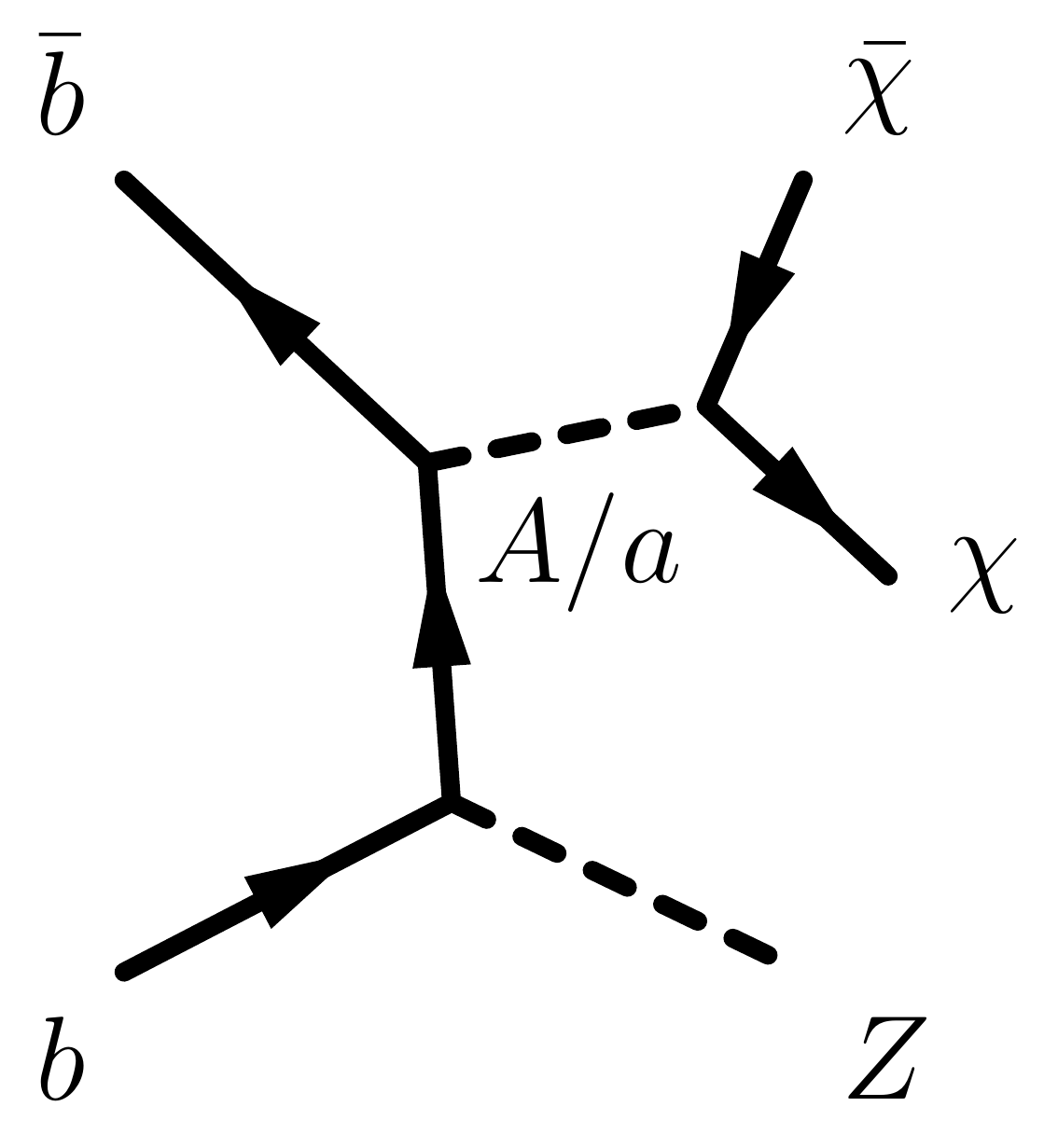}
} \\
\subfloat[]{\includegraphics[width=0.35\textwidth]{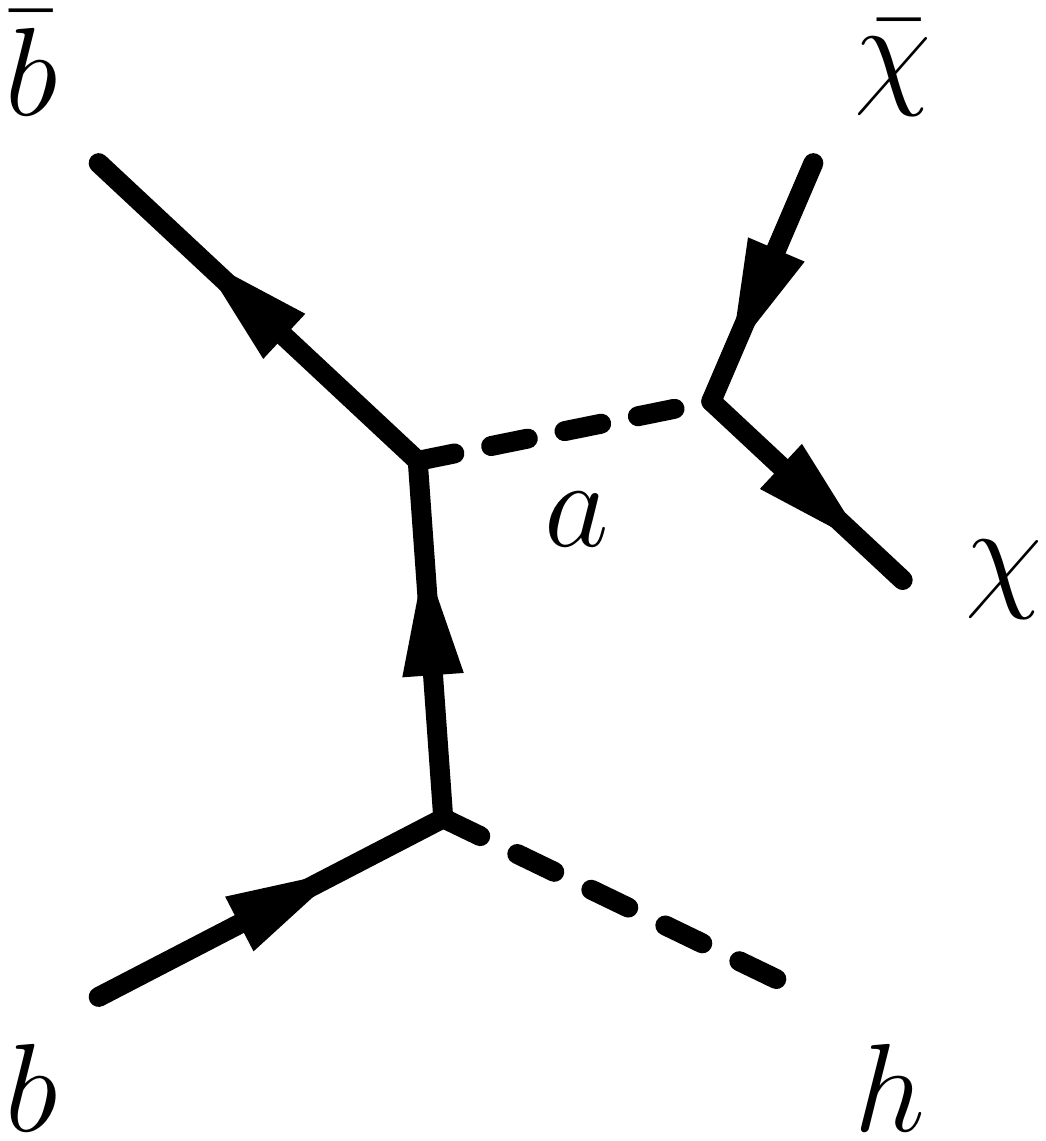}
}
\subfloat[]{\includegraphics[width=0.20\textwidth]{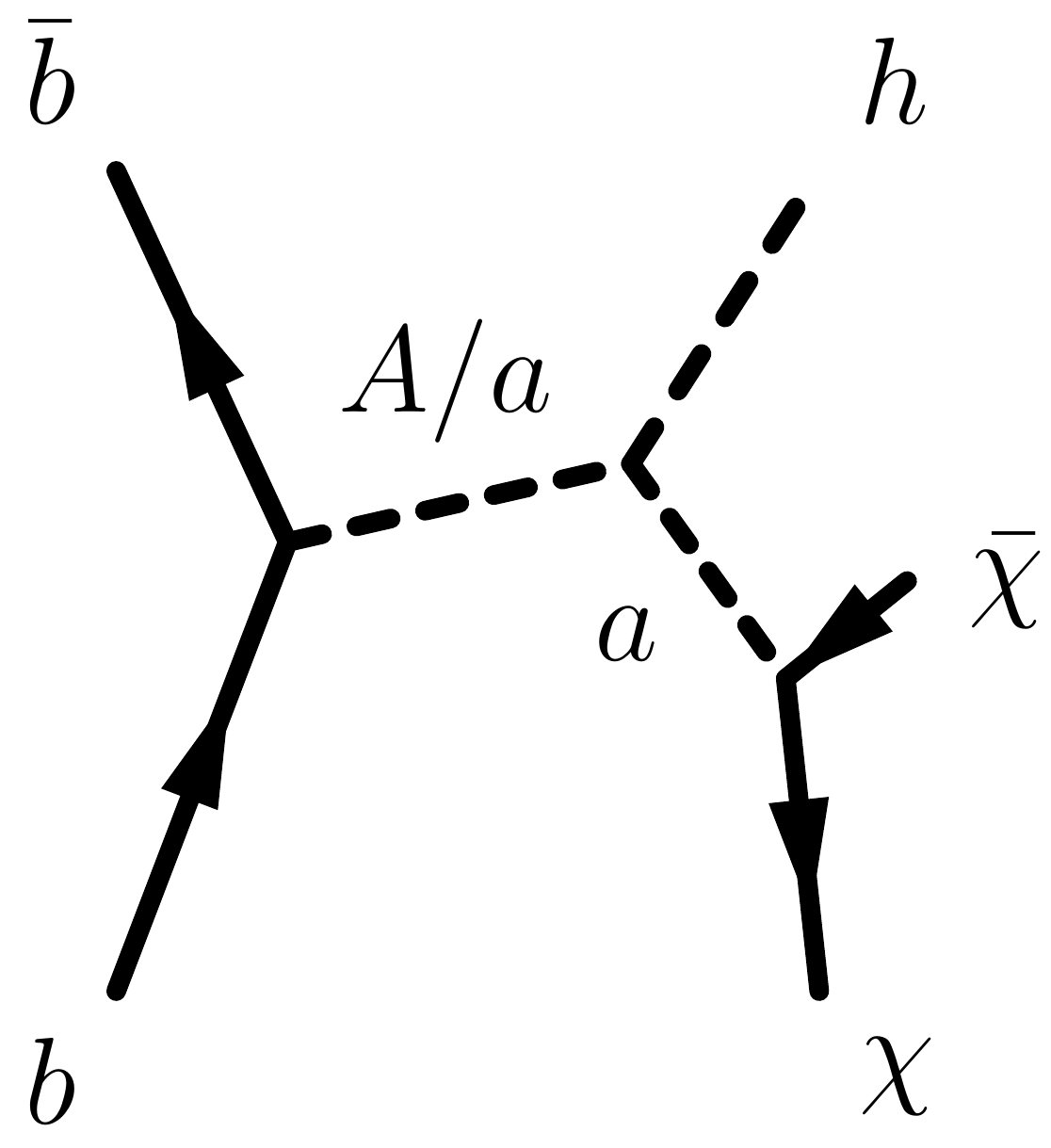}
}
\\
\subfloat[]{\includegraphics[width=0.21\textwidth]{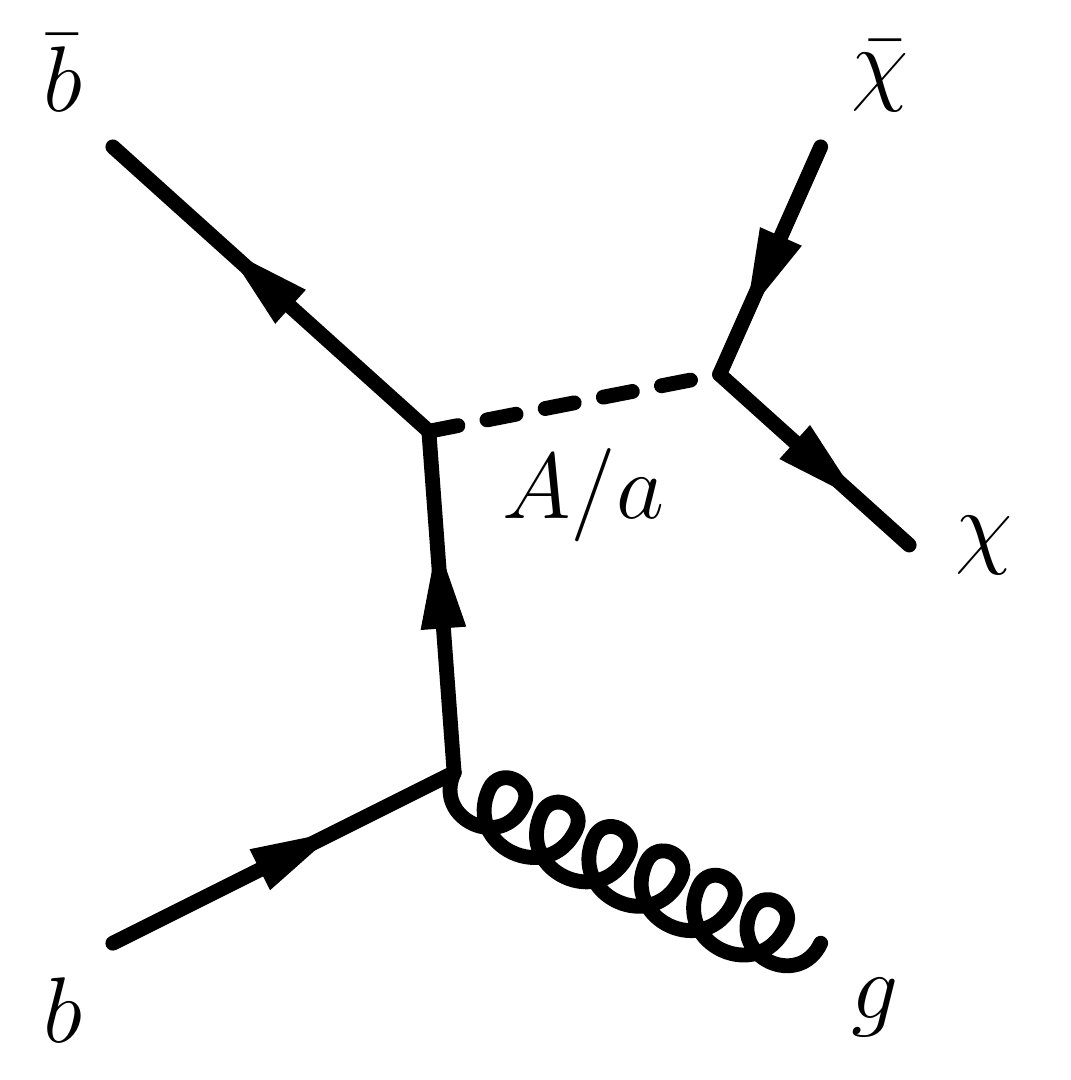}
\label{fig:THDMpa_feyn_monojet}
}
\caption{
\label{fig:THDMpa_feyn_bb}
Representative Feynman diagrams for the $b\bar{b}$-initiated production of
(a,b) the $\met+Z$ signature,
(c,d) the $\met+h$ signature,
and (e) the \met+$j$ signature in the \hdma.}
\end{figure}

\begin{figure}[h!]
\centering
\subfloat[]{\includegraphics[width=0.26\textwidth]{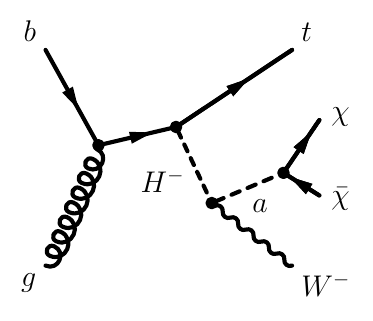}
}
\subfloat[]{\includegraphics[width=0.26\textwidth]{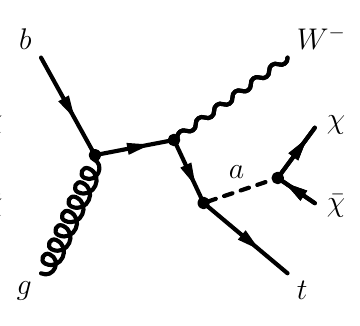}
} \\
\caption{
\label{fig:THDMpa_feyn_mettw}
Representative Feynman diagrams for the dominant production modes for the $\met+tW$ signature in the \hdma.}
\end{figure}

The \thdma has a particularly rich phenomenology, illustrated by the range of signatures shown in Figures~\ref{fig:THDMpa_feyn_gg}-~\ref{fig:THDMpa_feyn_mettw},
and this paper brings together an unprecedentedly large number of them.
The signatures can be group into those involving invisible and visible mediator decays, with the former being referred to as $\met+X$ signatures in the following.
At the LHC, the dominant production mode for the majority of signatures is $gg$-initiated production.
In Figure~\ref{fig:THDMpa_feyn_gg}, Feynman diagrams for the relevant signatures arising from $gg$-initiated production in the \thdma are summarised.
The $\met+Z$ and $\met+h$ signatures can be resonantly produced (Figure~\ref{fig:THDMpa_feyn_gg_res}), and non-resonantly (Figure~\ref{fig:THDMpa_feyn_gg_box}),
making them particularly relevant in the 2HDM+$a$ interpretation.
Additional signatures arising from $gg$-initiated production are the \met+$j$ signature (Figure~\ref{fig:THDMpa_feyn_gg_jet}),
resonant $A/H$ production with decay into \ttbar or \bbbar (Figure~\ref{fig:THDMpa_feyn_tt_res}),
\ttbar- or \bbbar-associated resonant $A/H$ production, leading to \tttt, $b\bar{b}b\bar{b}$, $t\bar{t}b\bar{b}$, $\met+\ttbar$, or $\met+\bbbar$ signatures (Figure~\ref{fig:THDMpa_feyn_res_assoc}),
$tb$-associated production of a charged Higgs boson decaying into $tb$, \htb (Figure~\ref{fig:THDMpa_feyn_htb}),
and production of a SM Higgs boson decaying into a pair of mediators $aa$ with subsequent decays into fermions or DM (Figure~\ref{fig:THDMpa_feyn_haa}).
Production from $b\bar{b}$ initial states for the $\met+Z$, $\met+h$, and $\met+j$ signatures is shown in Figure~\ref{fig:THDMpa_feyn_bb}.
Finally, the leading Feynman diagrams for the $\met+tW$ signature are shown in Figure~\ref{fig:THDMpa_feyn_mettw}.
The interplay between these signatures is highly dependent on the \thdma model parameters.

The phenomenology of the model is fully determined by 14 independent parameters:
the masses of the Higgs bosons $h$, $H$, $A$, and $H^{\pm}$;
the mass of the mediator $a$; the mass of the DM particle $\chi$;
the Yukawa coupling strength between the mediator and the DM particle, $g_{\chi}$;
the electroweak vacuum expectation value (VEV), $v$;
the ratio of the VEVs of the two Higgs doublets, $\tan\beta$;
the mixing angles of the CP-even and CP-odd weak eigenstates, $\alpha$ and $\theta$, respectively;
the quartic coupling $\lambda_{3}$ of the pure 2HDM potential term and the two quartic couplings of the potential terms connecting the doublet and singlet fields $\lambda_{\textrm{P1}}$ and $\lambda_{\textrm{P2}}$.
The values of some of these parameters are heavily constrained by both electroweak and flavour measurements and phenomenological considerations,
such as the requirement that the Higgs potential is stable~\cite{Bauer:2017ota,2HDMWGproxi}.
Further parameter choices are driven by the desire to simplify the phenomenology of the model and reduce the space of independent parameters to be scanned by experimental searches.
A summary of the parameter choices and the benchmark scenarios shown in this publication is given in the following.
A detailed description of the 2HDM+$a$ benchmark scenarios recommended by the LHC Dark Matter Working Group is given in Ref.~\cite{2HDMWGproxi}.

The following parameter settings are common to all benchmark scenarios described in Section~\ref{sec:summary}.
The coupling $g_{\chi}$ is set to unity with a negligible effect on the shapes of the kinematic distributions of interest.
The alignment and decoupling limits are assumed, hence $m_h = 125$~\GeV, $v=246$~\GeV, and $\cos(\beta-\alpha)=0$.
The quartic coupling $\lambda_3 = 3$ is chosen to ensure the stability of the Higgs potential for the choice of the masses of the heavy Higgs bosons.
The latter are fixed to the same value ($m_A=m_H=m_{H^{\pm}}$).
The choice $m_H = m_{H^{\pm}}$ is made to evade the constraints from electroweak precision measurements~\cite{Bauer:2017ota},
while the additional requirement $m_A = m_H$ is made to reduce the number of independent model parameters
~\cite{2HDMWGproxi}.\footnote{The mass differences $|m_A - m_H| \lesssim 200$~\GeV are consistent with constraints from electroweak precision measurements
and have the largest impact on the $\met+Z$ and $\met+h$ signatures due to the possibility of opening up the decay $H \rightarrow AZ$, which is not allowed in the mass-degenerate scenario
recommended by the LHC Dark Matter Working Group. For further discussions of scenarios with non-zero $|m_A - m_H|$ see Refs.~\cite{Bauer:2017ota,2HDMWGproxi}.}
The other quartic couplings are also set to 3 in maximise the trilinear couplings between the CP-odd and the CP-even neutral states.

After these considerations, five free parameters remain:
the mass of the heavy Higgs bosons, $m_A = m_H = m_{H^{\pm}}$;
the mass of the pseudo-scalar mediator, $m_a$;
the mass of the fermionic DM particle, $m_{\chi}$;
the sine of the mixing angle $\theta$ between the two CP-odd states $a$ and $A$, $\sin\theta$;
and the VEV ratio, $\tan\beta$.

The constraints on the model are evaluated for some representative benchmark scenarios,
in which one or two of the free parameters are varied while the others are kept at fixed values.
These benchmark scenarios, summarised in Table~\ref{tab:benchmarks}, are defined with the intention to highlight the diverse phenomenology of the \thdma
and to study the interplay and complementarities between different experimental signatures.

\textbf{Scenario 1: exploration of two $m_a$--$m_A$ planes.}
Constraints are evaluated as a function of the two pseudo-scalar masses \ma and \mA to highlight the complex dependence of the \thdma phenomenology
on the pseudo-scalar mass hierarchy, which determines the production and decay modes that are kinematically accessible and favoured.
The value of \tanb is fixed to 1.0, favouring couplings to up-type quarks, most notably top quarks.
Two choices of the $a-A$ mixing angle, \sint$=0.35$ and \sint$=0.7$, are explored.
These scenarios correspond to low and almost maximal mixing, respectively, between the pseudo-scalar $A$ belonging to the extended Higgs sector and the pseudo-scalar $a$ mediating the interaction with DM.

\textbf{Scenario 2: exploration of two $m_A$--$\tan\beta$ planes.}
The parameters \mA and \tanb are varied simultaneously for the two choices of the mixing angle \sint.
The pseudo-scalar mass is fixed to a value of 250~GeV, such that on-shell decays of the mediator into a pair of top quarks (\ttbar) are kinematically forbidden.
This means that the branching ratio for the invisible mediator decay $a\rightarrow \chi\chi$ can be as large as 100\%.
This benchmark scenario highlights the dependence of the couplings of the pseudo-scalar $A$ on the value of \tanb as a function of its mass.
Given the type-II Yukawa structure of the 2HDM+$a$, low values of \tanb correspond to a preferred coupling of $A$ (and $a$) to up-type quarks,
while higher values of \tanb imply stronger couplings to down-type quarks and charged leptons.
This benchmark scenario is evocative of the mass--$\tan\beta$ parameterisation used to summarise constraints on type-II 2HDMs,
such as the hMSSM, a Minimal Supersymmetric extension of the SM with a lighter scalar state at a mass of 125~\GeV~\cite{Djouadi:2013uqa}.
It also allows an exploration of the interplay between $gg$-initiated, top-loop induced and $b\bar{b}$-initiated production modes (see below).

\textbf{Scenario 3: exploration of two $m_a$--$\tan\beta$ planes.}
This scenario is similar to Scenario 2, with the difference that the mediator mass \ma is varied instead of the mass of the pseudo-scalar $m_A$,
which is fixed to a value of 600~\GeV. This means that decays of the pseudo-scalar $A$ into \ttbar are kinematically possible and favoured at low values of $\tan\beta$.
The choice of \mA is motivated by constraints on the mass of the charged Higgs boson ($m_{H^{\pm}}=m_A$) derived from precision measurements of $B$-meson decays~\cite{Misiak:2017bgg,Bauer:2017ota}.
Similarly to the previous scenarios, two choices of the $a-A$ mixing angle, \sint$=0.35$ and \sint$=0.7$, are studied.

\textbf{Scenario 4: variation of the mixing parameter $\sin\theta$.}
Constraints are evaluated as a function of the $a-A$ mixing parameter $\sin\theta$.
This benchmark scenario highlights the interplay between the $\met+X$ signatures, in particular $\met+Z$ and $\met+h$, which arise from invisible mediator decays,
and signatures that probe visible mediator decays.
This is due to the strong \sint dependence of the couplings $g_{Aha}$ ($g_{Aha}\propto \sint\cos\theta$) and $g_{HZa}$ ($g_{HZa}\propto \sin\theta$),
which affect $\met+h$ and $\met+Z$ production in the \thdma (see Figures~\ref{fig:THDMpa_feyn_gg_res} and~\ref{fig:THDMpa_feyn_gg_box}),
and the coupling $g_{a\ttbar}$, which plays a dominant role in the leading $\met+X$ production modes ($g_{a\ttbar}\propto \sin\theta$) (see Figure~\ref{fig:THDMpa_feyn_gg}).
As a consequence, for \sint$\rightarrow 0$, the sensitivity of the $\met+X$ signatures vanishes.

\textbf{Scenario 5: variation of the DM mass $m_{\chi}$.}
While the value of \mDM has a limited impact on the sensitivity of collider searches for $\mDM < \ma /~2$,
it has a strong effect on cosmological parameters, such as the relic density, and on the sensitivity of direct and indirect detection experiments.
This benchmark scenario therefore provides a basis for comparing the sensitivity of collider searches to
those of non-collider experiments and cosmological observations in the context of the 2HDM+$a$.
Only \mDM is varied, while the other free parameters are fixed to the following values: \sint$=0.35$, $m_A=600$~\GeV, $m_a=400$~\GeV, and $\tan\beta=1.0$.
The choice of the two mass parameters differs from that in the equivalent benchmark scenario described in Ref.~\cite{2HDMWGproxi} and explored in a previous ATLAS publication~\cite{EXOT-2017-32},
as the latter is fully excluded by the searches discussed in this publication.

\textbf{Scenario 6: exploration of a $m_a$--$m_\chi$ plane.}
This scenario serves to illustrate the interplay between searches for invisible and exotic decays of the light Higgs boson $h$ in the 2HDM+$a$.
Values of $\sin\theta=0.35$ and $\tan\beta=1.0$ are chosen for consistency with the other benchmark scans,
while a higher value $m_A = 1200$~\GeV is chosen to satisfy the constraint on the coupling $g_{haa}$
from measurements of the total Higgs boson decay width~\cite{Argyropoulos:2022ezr}. %
This is a powerful constraint on the low-$m_a$ region ($m_a < m_h /~2$), satisfied only by a relatively narrow range of $m_A$, for given values of the $\sin\theta$, $\tan\beta$ and quartic couplings $\lambda$.

In all benchmark scenarios other than Scenarios 5 and 6, $m_{\chi}=10$~\GeV is chosen.
This value ensures a sizeable branching ratio for the decay $a \rightarrow \chi\bar{\chi}$ for all values of $m_a > 100$~\GeV that are considered.
As shown in Section~\ref{sec:thdma_sum_mDM}, the choice of $m_{\chi}$ has a negligible impact on the sensitivity of the searches considered in this publication for $m_\chi < \ma/~2$.
Thus it is possible to match the observed relic density across a wide range of model parameter space through an appropriate choice of $m_\chi$, without impact on the experimental signatures.

In choosing the ranges for the parameters that are varied in a given benchmark scenario, various theoretical considerations are taken into account.
First, in some regions of the probed parameter space, the scalar potential is not bounded from below for large values of $m_A$.
For example, in Scenario 1a, this is the case for $m_A \gtrsim 1250$~\GeV ($m_A \gtrsim 1550$~\GeV) for $m_a = 100$~\GeV ($m_a = 1000$~\GeV).
However, these constraints can be relaxed substantially if the quartic couplings take a value closer to the perturbative limit
or in more general 2HDMs containing additional couplings as discussed in Refs.~\cite{2HDMWGproxi,Bauer:2017ota,Haisch:2018djm}.
Hence these should not be understood as strong limitations on the validity of the model predictions that were used to derive the exclusion contours.
Next, it is worth noting that, given these parameter choices, the $aah$ coupling exceeds the unitarity limit of $4\pi$ for large values of $m_A$.
For example, for the mentioned parameter choices $\sin\theta=0.35$ and $\tan\beta=1.0$ (Scenario 1a),
this is the case for $m_A \gtrsim 1250$~\GeV ($m_A \gtrsim 1500$~\GeV) for $m_a = 100$~\GeV ($m_a = 1000$~\GeV).
In this context, and for high $m_{A}$, the width of the additional heavy Higgs bosons grows substantially and the theoretical predictions
are subject to additional theoretical uncertainties from the treatment of the width.\footnote{The simulation of the signal processes considers effects
due to the off-shell production and decay of the Higgs bosons, but uses a fixed width to describe unstable resonances,
thus neglecting variation of the decay width as a function of the Higgs boson virtuality. This can have an impact outside of the resonance region.}
Therefore, regions where the relative width $\Gamma/m$ of at least one of the heavy Higgs bosons or that of the pseudo-scalar mediator exceeds 20\%
are marked as shaded areas in the summary figures in Section~\ref{sec:summary}.\footnote{These regions are mainly driven by the widths of the heavy Higgs bosons
rather than by the typically narrower width of the pseudo-scalar mediator.}
This is a conservative approach to indicate large widths and follows the choice in Ref.~\cite{EXOT-2017-32}.

\begin{table}[htbp]
\centering
\caption{Summary of the parameter settings for the different \thdma benchmark scenarios explored in this publication.}
\begin{tabular}{llcccccc}
\hline
\hline
\multicolumn{2}{l}{Scenario}   & \multicolumn{5}{c}{Fixed parameter values}  & Varied parameters \\
&        & \sint    &  \mA[\GeV] & \ma[\GeV] & \mchi[\GeV] & \tanb  &  \\

\hline
\multirow{2}{*}{1} & a &          $0.35$  &      --    &      --   & $10$ 	& $1.0$    &  \multirow{2}{*}{$(\ma ,m_A)$} \\
& b & $0.70$  &      --    &      --   & $10$       & $1.0$    & \\
\multirow{2}{*}{2} & a          & $0.35$  &      --    & $250$     & $10$       &     --        &  \multirow{2}{*}{$(m_A, \tan\beta)$} \\
& b & $0.70$  &      --    & $250$     & $10$       &      --        & \\
\multirow{2}{*}{3} & a          & $0.35$  & $600$      &      --   & $10$       &      --        & \multirow{2}{*}{$(m_a, \tan\beta)$} \\
& b & $0.70$  & $600$      &      --   & $10$       &      --        & \\
\multirow{2}{*}{4} & a          &     --  & $600$      & $200$     & $10$       & $1.0$  & \multirow{2}{*}{$\sin\theta$} \\
& b &     --  & $1000$     & $350$     & $10$       & $1.0$  & \\
5                           &   & $0.35$  & $1000$      & $400$    & --         & $1.0$  & \mDM \\
6                           &   & $0.35$  & $1200$     &      --   & --         & $1.0$  & ($m_a$, $m_\chi$) \\
\hline
\hline
\end{tabular}
\label{tab:benchmarks}
\end{table}

Scenarios 1a, 3a, 4a, 4b, and 5 are recommended by the LHC Dark Matter Working Group~\cite{2HDMWGproxi},
and were used in previous ATLAS publications, most notably Ref.~\cite{EXOT-2017-32}.
The additional scenarios, 1b, 2a, 2b, 3b, and 6 are motivated by the studies in Refs.~\cite{2HDMWGproxi,Pani:2017qyd,Argyropoulos:2022ezr}.
In particular, the choice of $\sin\theta=0.7 \approx 1/\sqrt{2}$ ($\theta = \pi/4$) corresponds to maximal mixing in the pseudo-scalar sector
and is particularly relevant, for example, for the $\met+tW$ search, which was designed specifically for 2HDM+$a$ signal processes~\cite{Pani:2017qyd}.
Scenario 6 is shown for the first time in this publication to highlight further the rich phenomenology of the model.

Another improvement introduced in this publication concerns the production modes of the various Higgs bosons and the pseudo-scalar mediator.
In the previous comprehensive summary publication of ATLAS DM searches~\cite{EXOT-2017-32}, only $gg$-initiated production was considered for the $\met+Z$ signatures.
For the $\met+h$ signatures, $b\bar{b}$-initiated production was taken into account but only for values of $\tanb>10$.
In this publication, $b\bar{b}$-initiated production is taken into account for all $\met+X$ signatures,
which is particularly relevant for the $\met+Z$ and $\met+h$ signatures at large values of $\tan\beta$, but also contributes at intermediate values.
%
%

% End of text imported from the .//sections/thdma.tex input file

%
\section{ATLAS detector}
\label{sec:detector}
%

% The next lines are included from the .//sections/atlas-detector.tex input file
The ATLAS detector~\cite{PERF-2007-01} is a multipurpose particle detector with a forward--backward symmetric cylindrical
geometry and a near 4$\pi$ coverage in solid angle.\footnote{
ATLAS uses a right-handed coordinate system with its origin at the nominal interaction point in the centre of the detector.
The positive $x$-axis is defined by the direction from the interaction point to the centre of the LHC ring,
with the positive $y$-axis pointing upwards, while the beam direction defines the $z$-axis.
Cylindrical coordinates $(r,\phi)$ are used in the transverse plane, $\phi$ being the azimuthal angle around the $z$-axis.
The pseudorapidity $\eta$ is defined in terms of the polar angle $\theta$ by $\eta=-\ln\tan(\theta/2)$,
while the rapidity $y$ is defined as $y = 0.5 \ln[(E + p_z)/(E-p_z)]$, where $E$ denotes the energy and $p_z$ the component of the momentum along the beam direction.
The angular distance $\Delta R$ is defined as $\sqrt{(\Delta \eta)^2 + (\Delta \phi)^2}$.
}
It consists of an inner tracking detector surrounded by a thin superconducting solenoid providing a \qty{2}{\tesla} axial magnetic field,
electromagnetic and hadron calorimeters, and a muon spectrometer.
The inner-detector system (ID) is immersed in a \qty{2}{\tesla} axial magnetic field and provides charged-particle tracking in the range of \(|\eta| < 2.5\).
The high-granularity silicon pixel detector covers the vertex region and typically provides four measurements per track,
the first hit normally being in the insertable B-layer (IBL) installed before Run~2~\cite{ATLAS-TDR-19,PIX-2018-001}.
It is followed by the silicon microstrip tracker (SCT), which usually provides eight measurements per track.
These silicon detectors are complemented by the transition radiation tracker (TRT), which enables radially extended track reconstruction up to \(|\eta| = 2.0\).
The TRT also provides electron identification information based on the fraction of hits (typically 30 in total) above a higher energy-deposit threshold corresponding to transition radiation.
The calorimeter system covers the pseudorapidity range \(|\eta| < 4.9\).
Within the region \(|\eta|< 3.2\), electromagnetic (EM) calorimetry is provided by barrel and endcap high-granularity lead/liquid-argon (LAr) calorimeters,
with an additional thin LAr presampler covering \(|\eta| < 1.8\) to correct for energy loss in material upstream of the calorimeters.
Hadron calorimetry is provided by the steel/scintillator-tile calorimeter, segmented into three barrel structures within \(|\eta| = 1.7\), and two copper/LAr hadron endcap calorimeters.
The solid angle coverage is completed with forward copper/LAr and tungsten/LAr calorimeter modules optimised for electromagnetic and hadronic energy measurements respectively.
The muon spectrometer (MS) comprises separate trigger and high-precision tracking chambers measuring the deflection of muons in a magnetic field generated
by the superconducting air-core toroidal magnets.
The field integral of the toroids ranges between \num{2.0} and \qty{6.0}{\tesla\metre} across most of the detector.
Three layers of precision chambers, each consisting of layers of monitored drift tubes, cover the region \(|\eta| < 2.7\),
complemented by cathode-strip chambers in the forward region, where the background is highest.
The muon trigger system covers the range \(|\eta| < 2.4\) with resistive-plate chambers in the barrel, and thin-gap chambers in the endcap regions.
The ATLAS trigger system consists of a first-level trigger system implemented in custom hardware
followed by a software-based high-level trigger~\cite{TRIG-2016-01}.
The level-1 trigger uses a subset of the detector information to accept events at a rate below \qty{100}{ kHz},
while the software-based trigger reduces the accepted event rate to \qty{1}{ kHz} on average depending on the data-taking conditions.
An extensive software suite~\cite{ATL-SOFT-PUB-2021-001,} is used in the reconstruction and analysis of real and simulated data,
in detector operations, and in the trigger and data acquisition systems of the experiment.

% End of text imported from the .//sections/atlas-detector.tex input file

%
\section{Data and simulated event samples}
\label{sec:samples}
%

% The next lines are included from the .//sections/samples.tex input file
All analyses discussed in this publication are based on data from proton--proton collisions at a centre-of-mass energy of $\sqrt{s}=13$~\TeV collected with the ATLAS detector
at the LHC in the years 2015--2018, unless otherwise stated.
The integrated luminosity of the data sample, after requiring that all detector subsystems were operational and recording good quality data~\cite{DAPR-2018-01}, is 139~\ifb.

Simulated data are used to model the background processes and the predictions of the \thdma benchmark.
Details of the Monte Carlo (MC) generation for the various background processes considered in the analyses interpreted in this publication
are found in the individual analysis publications referenced in Section~\ref{sec:analyses}.
The \thdma benchmark is implemented in the Universal FeynRules Output (UFO) format~\cite{Degrande:2011ua}.
The implementation is referred to as \texttt{Pseudoscalar\_2HDM} in the following.
All signal processes, with the exception of the \htb process (see Table~\ref{tab:MCscalarsignal}),
are generated at leading-order (LO) in the strong coupling constant,
where LO means loop-induced gluon-gluon fusion for the $\met+X$ signatures (Figure~\ref{fig:THDMpa_feyn_gg}).

Events were generated from this UFO implementation using the \MGNLO~\cite{Alwall:2014hca} MC generator
interfaced with \PYTHIA~8~\cite{Sjostrand:2014zea} for the modelling of the parton shower and hadronisation
with the parameter values set according to the ATLAS tune A14~\cite{ATL-PHYS-PUB-2014-021}.
\MGNLO versions ranging from 2.6.0 to 2.9.5 and \PYTHIA versions ranging from 8.212 to 8.245 were used, depending on the analysis,
as summarised in Table~\ref{tab:MCscalarsignal}.
No differences between the signal simulations are expected to arise from the different choices of generator versions.
The NNPDF3.0NLO~\cite{Ball:2014uwa} set of parton distribution functions (PDF) at next-to-leading-order in the five-flavour scheme was used,
which assumes a massless $b$-quark and $\alpha_{s}(m_{Z}) = 0.118$~\cite{Ball:2014uwa}.
For consistency, the five-flavour scheme and $m_b=0~\GeV$ were chosen for the matrix element (ME) computation in \MGNLO for the $b\bar{b}$-initiated production.
For the $gg$-initiated production the four-flavour scheme was used to include top and bottom quark contributions in the production loop.
These modelling choices follow the recommendations of the LHC Dark Matter Working Group~\cite{2HDMWGproxi}.

To simulate the effects of additional $pp$ collisions in the same and nearby bunch crossings,
additional interactions were simulated using the soft QCD processes of \PYTHIA~8.186 with the A3 tune~\cite{ATL-PHYS-PUB-2012-003} and the MSTW2008LO PDF~\cite{Martin:2009iq},
and overlaid onto each simulated hard-scatter event.
The simulated samples were reweighted to reproduce the instantaneous luminosity spectrum in the data.
The simulations include the expected bunch train structure in data and include low-level corrections to account for bunch train effects.
Simulated events were processed either through a detector simulation~\cite{SOFT-2010-01} based on \GEANT~\cite{Agostinelli:2002hh} or
through a fast simulation~\cite{ATL-PHYS-PUB-2010-013} with a parameterisation of the calorimeter response and \GEANT for the other parts of the detector.
All simulated samples were reconstructed in the same manner as the data. Corrections derived from data control samples were applied to simulated events
to account for differences between data and simulation in the reconstruction efficiencies, the energy/momentum scale and resolution of reconstructed electrons and muons,
and in the efficiency and false positive rate for identifying jets containing $b$-hadrons.
The energy scale and resolution of hadronic jets are also corrected to give the same performance between data and MC.

To produce signal events efficiently across the large multi-dimensional parameter space of the 2HDM+$a$,
the \mg\ reweighting module~\cite{Mattelaer:2016gcx} was used to obtain predictions for a range of different signal model parameters from a minimal set of generated events.
This was achieved by assigning new event weights based on the ratios of matrix-elements for the input (generated) and target parameter points.
The event weights were calculated on-the-fly during the event simulation.
This method was validated by comparing weighted distributions with generated ones for a few representative samples.
The reweighting immensely reduces of the required computing resources as the detector simulation need be run only once.

\begin{table}
\caption{Details of the \MGNLO generation set-up used for the \thdma signal samples, for the signatures considered in this publication.
The \texttt{Pseudoscalar\_2HDM} UFO model is used for all simulated samples except those for the \htb search, which relies on the UFO of Ref.~\cite{Degrande:2015vpa}.
The \hinv\ and \hlrs signatures are not listed here as no signal samples required for the re-interpretation, which in those cases relies on the branching ratio limits,
as explained in Sections~\ref{subsubsec:hinv} and~\ref{subsubsec:hlrs}, respectively.}
\label{tab:MCscalarsignal}
\centering
\scalebox{0.75}{
\begin{tabular}{m{26mm}m{90mm}m{16mm}m{26mm}}
\hline\hline
Analysis & Generator and Parton Shower & Cross-section & Further details \\
\toprule
\monozll  & \MGNLO 2.4.3 (LO) + \PYTHIA 8.212 & LO & \\
\monohbb  & \MGNLO 2.6.0 (LO) + \PYTHIA 8.212\tablefootnote{With the exception of the $\ma-\mDM$ scan, where \MGNLO 2.7.4 (LO) + \PYTHIA 8.244 is used.} & LO & \\
\monohgamgam  & \MGNLO 2.7.3 (LO) + \PYTHIA 8.244& LO & \\
\monohtautau  & \MGNLO 2.7.3 (LO) + \PYTHIA 8.244& LO & \\
$\met+j$  & \MGNLO 2.7.3 (LO) + \PYTHIA 8.244 & LO & Section~\ref{subsubsec:metj} \\
$\met+tW$ & \MGNLO 2.7.3 (LO) + \PYTHIA 8.244 & LO &  \\
\tttt     & \MGNLO 2.9.5 (LO) + \PYTHIA 8.245 & LO & Ref.~\cite{Bauer:2017ota} \\
\htb      & \MGNLO 2.2.2 (NLO) + \PYTHIA 8.212 & NLO, 4FS & Section~\ref{subsubsec:htb}  \\
\hline
\hline
\end{tabular}
}
\end{table}

\FloatBarrier

% End of text imported from the .//sections/samples.tex input file

%
\section{Experimental signatures}
\label{sec:analyses}
%

% The next lines are included from the .//sections/input_analyses.tex input file
\FloatBarrier

% The next lines are included from the .//sections/input_analyses_summary.tex input file
A wide range of searches in different final states targeting invisible or visible mediator decays probe the 2HDM+$a$.
No significant deviation from the SM prediction was observed in any of these searches,
hence they are used to derive constraints on the \thdma for benchmark scenarios introduced in Section~\ref{sec:thdma}.
The sensitivity of searches varies across different regions of the \thdma parameter range
and not all searches are therefore interpreted in all \thdma benchmark scenarios.
In Table~\ref{tab:input_summary}, an overview of the searches interpreted in the context of different \thdma benchmark scenarios is given.
The individual searches are summarised in the following subsections.
Further details can be found in the individual publications referenced at the beginning of each subsection.
The \monohbb, \monozll, and \htb searches enter the statistical combination described in Section~\ref{sec:comb}.

The analyses rely on final-state physics objects that are reconstructed using information from the different subsystems of the ATLAS detector.
Jets are reconstructed from particle-flow objects~\cite{PERF-2015-09}
using the anti-$k_t$ algorithm~\cite{Cacciari:2008gp,Fastjet} with radius parameter $R=0.4$ (small-$R$ jets) and $R=1.0$ (large-$R$ jets)~\cite{JETM-2018-05}.
Multivariate algorithms are used to identify small-$R$ jets within $|\eta|=2.5$ containing $b$-hadrons ($b$-jets)~\cite{FTAG-2019-07,FTAG-2018-01}.
Photons are reconstructed from topologically connected clusters of energy deposits in the EM calorimeters~\cite{EGAM-2018-01}.
Electrons are reconstructed from topologically connected energy clusters~\cite{ATL-PHYS-PUB-2017-022,}
in the EM calorimeters matched to a charged-particle track in the ID~\cite{EGAM-2018-01}.
Muons are reconstructed from matching tracks in the ID and MS, refined through a global fit which uses the hits from both the subdetectors~\cite{PERF-2015-10}.
The analyses may implement different lepton and photon selection criteria for
particle identification, isolation, and kinematic requirements, for example \pT\ and $\eta$.
The reconstruction of $\tau$-leptons depends on the $\tau$-lepton decay (hadronic or leptonic) targeted by a given analysis.
The visible part of hadronically decaying $\tau$-leptons~\cite{ATL-PHYS-PUB-2015-045,ATLAS-CONF-2017-029} is seeded by
small-$R$ jets reconstructed from topological clusters, calibrated with a hadronic weighting scheme~\cite{Barillari:1112035}.
The missing transverse momentum $\vec{p}_{\mathrm{T}}^{\;\mathrm{miss}}$ (with magnitude $\met$) is calculated from the negative vector sum of transverse momenta ($\pt$) of electrons, muons
and jet candidates and an additional soft term~\cite{PERF-2016-07} which includes activity in the tracking system originating from the primary vertex
but not matched with any reconstructed particle.
Some analyses may also consider photons and $\tau$-leptons in the \met\ reconstruction.

\begin{table}
\caption{Summary of input analyses used in the different benchmark scenarios.}
\label{tab:input_summary}
\centering
\begin{tabular}{lccccccccccc}
\hline
\hline
Analysis/Scenario  & 1a & 1b & 2a & 2b & 3a & 3b & 4a & 4b & 5 & 6 \\
\midrule
\monozll~\cite{HIGG-2018-26} & x & x & x  & x  & x & x  & x  & x  & x  &     \\
\monohbb~\cite{EXOT-2018-46} & x & x & x  & x  & x & x  & x  & x  & x  & x   \\
\monohgamgam~\cite{HIGG-2019-02} & x & x &   &   & x & x & x & x &  &    \\
\monohtautau~\cite{HDBS-2018-50} & x &   &   & x &   &   &   &   &   &   \\
$\met+tW$~\cite{EXOT-2021-01} & x & x & x & x & x  & x  & x  & x  &   &   \\
$\met+j$~\cite{EXOT-2018-06} & x & x &   &  & x & x & x  & x  &   &   \\
\hinv~\cite{HIGG-2021-05} & x & x &   &   & x &   &   &   &   & x   \\
$\met+Z(q\bar{q})$~\cite{EXOT-2016-23} & x &  &   &   &  &   & x  & x  &    &     \\
$\met+\bbar$~\cite{SUSY-2016-18} &   &   &    &    &   &    & x  & x  &    &     \\
$\met+\ttbar$~\cite{SUSY-2016-18,SUSY-2016-16} &  &   &    &    &   &   & x  & x  &    &     \\
\tttt~\cite{EXOT-2019-26} & x & x & x & x & x & x & x & x & x &     \\
\htb~\cite{HDBS-2018-51}  & x & x & x & x & x & x & x & x & x &     \\
\hlrs~\cite{HDBS-2021-03,HIGG-2017-05,HIGG-2014-02,EXOT-2016-22,HDBS-2018-55}  &   &   &   &   &   &   &   &   &   & x   \\
\hline
\hline
\end{tabular}
\end{table}

% End of text imported from the .//sections/input_analyses_summary.tex input file

\subsection{Searches for invisible mediator decays}
%

% The next lines are included from the .//sections/input_analyses_metzll.tex input file
\subsubsection{\monozll}

Signal events in this analysis~\cite{HIGG-2018-26} are required to have \met and a pair of high-\pT leptons ($\ell = e, \mu$).
They are required to satisfy a set of single-electron~\cite{TRIG-2018-05} or single-muon~\cite{TRIG-2018-01} triggers
which require the presence of an electron (muon) with transverse energy (transverse momentum) above thresholds in the range of 20–26 GeV
depending on the lepton flavour and data-taking period~\cite{TRIG-2019-03}.
Accordingly, a requirement of $\pT > 30$~\GeV is imposed on the leading electron or muon in the event,
while the subleading lepton is required to satisfy $\pT > 20$~\GeV.
The leptons are required to have the same flavour, be oppositely charged, and their invariant mass must be between 76~\GeV and 106~\GeV,
compatible with the $Z$ boson mass.
To select events consistent with invisible particles recoiling against the $Z$ boson, events are required to have $\met > 90$~\GeV and $S_{\met} > 9$,
where $S_{\met}$ denotes the object-based \met significance~\cite{ATLAS-CONF-2018-038}.
Additionally, a requirement $\Delta R(\ell\ell) < 1.8$ on the angular separation between the two leptons is required.
Events with one or more $b$-jets are removed in all regions to suppress events containing top quarks.

The dominant background is the $ZZ$ background, followed by $WZ$, $Z$+jets, and the non-resonant backgrounds ($WW$, \ttbar, single top-quark, and $Z\rightarrow\tau\tau$).
Additional smaller contributions arise from triboson production, $\ttbar+V$, and $ZZ \rightarrow 4\ell$, where two of the leptons are not reconstructed.
The backgrounds from $ZZ$ and $WZ$ production and the sum of the non-resonant backgrounds are estimated from MC simulation
and normalised to data in the final likelihood fit using dedicated $4\ell$, $3\ell$, and $e\mu$ control regions, which are enriched in the respective background components.
The remaining, smaller, backgrounds are estimated from MC simulation.
The final analysis result is obtained from a simultaneous profile likelihood fit in the $ee$ and $\mu\mu$ signal and the $4\ell$, $3\ell$, and $e\mu$ control regions.
The observable of interest in the signal regions and the $e\mu$ control regions is the transverse mass
\begin{equation}\nonumber
\mt=\sqrt{\left[\sqrt{m_Z^2+(\pt^{\ell\ell})^2}+\sqrt{m_Z^2+(\met)^2}\right]^2-\left[\vec{p}_\mathrm{T}^{\ell\ell}+\vec{p}_{\mathrm{T}}^{\;\mathrm{miss}}\right]^2},
\end{equation}
which provides a good separation between the \thdma signal and the dominant $ZZ$ background.
Only events with $\mt>200$~\GeV are included in the final fit. In the $4\ell$ and $3\ell$ control regions, the \met distribution is fitted.

% End of text imported from the .//sections/input_analyses_metzll.tex input file

% The next lines are included from the .//sections/input_analyses_methbb.tex input file
\subsubsection{\monohbb}

The \monohbb\ analysis signature consists of two $b$-jets and significant \met\ coming from the decays of a SM Higgs boson
and a light pseudo-scalar to dark matter respectively~\cite{EXOT-2018-46}.
Events are required to pass the \met\ trigger~\cite{TRIG-2019-01} and to have $\met>150$~GeV, with at least two jets identified as $b$-jets.
Selections split the events into categories with exactly two and greater than two $b$-jets,
to give good sensitivity to both the gluon--gluon fusion and $b\bar{b}$-initiated production processes,
which are significant at low and high values of $\tan\beta$.
The Higgs boson recoils against the pseudo-scalar which decays into dark matter in the signal topology,
hence the \met\ and Higgs-boson \pt are strongly correlated.
For this reason, a $\met<500$~\GeV\ requirement is used to separate the resolved topology,
in which the $b$-jets are reconstructed as two separate small-$R$ jets, from the merged one,
in which the high momentum of the Higgs boson implies that both the $b$-quarks can be found within a single large-$R$ jet.
Both the topologies are further subdivided into few \met\ ranges.
The analysis is performed through a simultaneous fit of the observed $m_{bb}$ distribution across all signal regions and the yields of the control regions.

The dominant backgrounds arise from \ttbar and $Z/W$-boson production with jets containing heavy flavour quarks.
Smaller contributions from single-top, diboson and SM $Vh$ production are also present.
SM processes generating \met\ through the leptonic decay of a $W$ boson are reduced by rejecting events containing electron or muons.
The contribution from $Z+$jets processes becomes increasingly dominant for high \met\ selections.
In the resolved category, events must also satisfy a requirement $S_{\met}>12$ which suppresses the multijet background to negligible levels.
Additional requirements are made on the transverse mass of the \met\ and the $b$-jets to reduce contamination from \ttbar processes.
Finally, requirements are made on the reconstructed \pT of the Higgs boson candidate, which increases with \met,
and on the number of additional jets in the event. Both of these also serve to reduce background contributions.
Control regions requiring one or two leptons are used to normalise and validate the simulations used to model the main background processes for each signal selection.

% End of text imported from the .//sections/input_analyses_methbb.tex input file

% The next lines are included from the .//sections/input_analyses_methyy.tex input file
\subsubsection{\monohgamgam}

The \monohgamgam\ analysis targets final states with two photons and significant \met~\cite{HIGG-2019-02}.
Events are selected using a diphoton trigger requiring two reconstructed photon candidates
with minimum transverse energies of 35 and 25~\GeV for the leading and subleading photons, respectively.
Events are required to contain at least two photon candidates and $\met > 90$~\GeV.
The two photons with highest energy in the transverse plane are selected to form a Higgs boson candidate
if they satisfy the requirements $E_\mathrm{T}^{\gamma}/m_{\gamma\gamma} > 0.35$ and 0.25, respectively, where $m_{\gamma\gamma}$ is the invariant mass of the two selected photons.
Furthermore, events are required to have 105~\GeV $< m_{\gamma\gamma} < 160$~\GeV.
The data sideband is defined to use events in this region but excluding the region 120~\GeV $< m_{\gamma\gamma} < 130$~\GeV.
Following this preselection, a boosted decision tree (BDT) is trained to discriminate between the \thdma signal and the non-resonant diphoton backgrounds,
using variables such as \pTgg and $S_{\met}$ as inputs.
Finally, events are separated into low \etmiss ($\etmiss < 150$~\GeV) and high \etmiss ($\etmiss > 150$~\GeV) regions.i
In each region, two categories are defined from two sequential ranges of the BDT score,
with the ranges optimised to maximise the combined signal sensitivity in the two chosen categories while discarding the remaining events.

The main backgrounds arise from SM Higgs boson production, QCD-induced non-resonant diphoton production ($\gamma\gamma$ and $V\gamma\gamma$, where $V$ is a $W$ or $Z$ boson),
and reducible contributions where an electron or a jet is mis-identified as a photon and \met is generated either by particles escaping the detector acceptance
or by neutrinos ($V\gamma$, $\gamma$+jet). An additional background contribution dominating the low \met region originates from resolution effects
when computing the transverse energy from high-energy objects and softer contributions measured in the ID.
The background contributions are estimated by fitting analytic functions to the diphoton invariant mass distribution in the range of $105<m_{\gamma\gamma}<160$~\GeV
in each of the four signal-region categories.

%
%

% End of text imported from the .//sections/input_analyses_methyy.tex input file

% The next lines are included from the .//sections/input_analyses_methtautau.tex input file
\subsubsection{\monohtautau}

The \monohtautau search targets dark matter produced in association with a Higgs boson in final states
with two hadronically decaying $\tau$-leptons and missing transverse momentum~\cite{HDBS-2018-50}.
It is optimised specifically to search for the 2HDM+$a$.
Events are required to satisfy a combined di-$\tauhadvis$ + \met\ trigger~\cite{TRIG-2016-01,ATLAS-CONF-2017-061},
where $\tauhadvis$ denotes the visible part of a hadronically decaying $\tau$-lepton.
They are required to contain exactly two $\tau$-lepton objects that geometrically match the trigger-level $\tau$-lepton candidates
activating the di-$\tau$-lepton+\met trigger.
The leading $\tau$-lepton is required to have $\pT > 40$--$65$~\GeV, depending on the trigger threshold of a given data-taking year,
while the sub-leading $\tau$-lepton is required to have $\pT > 30$~\GeV.
The events also must satisfy $\met > 150$~\GeV to ensure that the trigger is operating at maximum efficiency.
Events containing an electron or a muon are vetoed. Events are further required to have at most one $b$-jet.
In addition to these pre-selection requirements, two non-orthogonal signal regions are constructed to target signal model parameter configurations
with high and low masses of the heavy pseudo-scalar $A$, respectively.
The signal regions with the stronger expected exclusion for a given signal hypothesis is used to derive the exclusion limits for this hypothesis.
Each signal region is further subdivided using the sum of the transverse masses of the two $\tau$-leptons, $m^{\tau_1}_{\text{T}}+m^{\tau_2}_{\text{T}}$,
where:
\begin{equation} \nonumber
m_\text{T}^{\tau_{i}} = \sqrt{2\pt^{\tau_{i}} \met(1 - \cos \Delta \phi(\tau_{i},\pt^\text{miss}))}.
\end{equation}
The requirement $m^{\tau_1}_{\text{T}}+m^{\tau_2}_{\text{T}} > 100$~\GeV is imposed to suppress events from $Z(\tau\tau)$+jets production,
in which the \met vector is typically collinear with the di-$\tau$-lepton system.

Higgs boson production in association with a $Z$ boson decaying into neutrinos is an irreducible background in this search.
Further background contributions arise from $Z$+jets, $VV$, \ttbar, multijet, and $Wh$ production.
SM background processes are modelled using a combination of simulated events and data-driven methods.
Background processes with only true $\tau$-leptons, mostly $Z$+jets, $VV$, and $Vh$ production and most of the \ttbar background,
are modelled using simulation normalised to the data in the dedicated control regions.
Events with at least one fake $\tau$-lepton, i.e. a non-$\tau$-lepton object mis-identified as a $\tau$-lepton, are estimated by using data-driven techniques.

% End of text imported from the .//sections/input_analyses_methtautau.tex input file

% The next lines are included from the .//sections/input_analyses_mettw.tex input file
\subsubsection{$\met+tW$}

The search considers final states with zero or one charged lepton ($\ell = e, \mu$), at least one $b$-jet and large missing transverse momentum~\cite{EXOT-2021-01}.
In addition, a result from a previous search~\cite{EXOT-2018-43} considering final states with two charged leptons is included in the interpretation of the results.
The signal regions for the zero- and one-lepton channels in Ref.~\cite{EXOT-2021-01} are designed to be orthogonal to each other and to the signal region
of for the two-lepton channel in Ref.~\cite{EXOT-2018-43} and are statistically combined in the final fit.

The search is optimised specifically for signals arising in the context of the \thdma
and is particularly sensitive to on-shell production of the charged Higgs bosons $H^{\pm}$ and their semi-invisible decays via the mediator particle, $a$:
$H^{\pm}\rightarrow W^{\pm}a(\chi\bar{\chi})$.
Due to the similarity of the experimental signature to $\ttbar$ production, the analysis is also sensitive to DM produced in association with two top quarks ($\met+\ttbar$).
This final state is not considered in the optimisation of the analysis regions but its contribution is added to the $\met+tW$ signal,
according to the prediction of the 2HDM+$a$, in the interpretation of the final result.
Candidate events were recorded using a combined set of triggers %
based on the presence of missing transverse momentum or charged leptons
and are required to have $\met>250$~\GeV ($\met>200$~\GeV for the two-lepton channel).
Further event selection criteria differ between analysis channels and are defined based on the number and type of leptons, jets and $b$-jets,
and a number of event variables, such as invariant and transverse masses and the angular separation between selected objects.

The relative importance of SM background processes varies across the different signal regions, although the main sources can be broadly classified
by either the presence of genuine \met produced by neutrinos, or false \met signals due to the mis-identification of particles, mis-measurements of their properties,
due to particles outside the kinematic acceptance of the detector, or pile-up.
Examples of the former include the $Z$+jets background in the zero-lepton channel and the $W$+jets background in the one-lepton channel.
SM \ttbar production and $W$+jets production in the zero-lepton channel are examples of major backgrounds with false \met signals due to leptons
that are either outside of the detector significance or mis-identified as jets.
Further backgrounds include those from $\ttbar Z$ and single top-quark production.
All background components are estimated from MC simulation.
Dedicated control regions are used to constrain the normalisation parameters of the five dominant background components in the final likelihood fit.

% End of text imported from the .//sections/input_analyses_mettw.tex input file

% The next lines are included from the .//sections/input_analyses_metj.tex input file
\subsubsection{\monojet}
\label{subsubsec:metj}

This search targets production of a single jet with large \met~\cite{EXOT-2018-06}.
The data was collected using the \met\ trigger.
Events are required to have $\met > 200$~\GeV to ensure that the trigger is fully efficient for events passing the analysis selection criteria.
They are also required to contain at least one jet with $\pT > 150$~\GeV with $|\eta|<2.4$, up to three additional jets with $\pT > 30$~\GeV and $|\eta|<2.8$,
and no reconstructed leptons ($e$, $\mu$ or $\tau$-leptons) or photons.
Several signal regions are considered with increasing requirements on the missing transverse momentum starting at 200~\GeV.
Additional angular requirements on the separation in $\phi$ between the \met vector and leading jet are imposed to reduce the contribution
from multijet events with mis-measured jet energies.

The dominant SM background for this search arises from $Z(\nu\nu)$ and $W(\ell\nu)$ production with jets,
where the $W$ boson decays into either hadronically decaying $\tau$-leptons or undetected electrons or muons.
Additional background contributions include \ttbar and single-top production, diboson production,
as well as non-collision and multijet backgrounds.
The estimate of the major SM processes in the analysis selection is based on a profile likelihood fit
to the distribution of the \pT of the system recoiling against the jets reconstructed in the event,
performed simultaneously in the signal region and in orthogonal control regions enriched with the targeted backgrounds.

Various different signal contributions to the $\met+j$ signal regions are considered in the re-interpretation of this search in the context of the 2HDM+$a$.
Production of a pair of DM particles with a jet in the matrix element ($pp \rightarrow \chi\chi j$) is the dominant signal contribution
in the signal regions of the \monojet\ analysis for low values of \met ($\met \lesssim 500$~\GeV) if the mediator mass is not too small ($\ma \gtrsim 150$~\GeV).
Both the loop-induced, $gg$-initiated (Figure~\ref{fig:THDMpa_feyn_gg_jet}) and tree-level, $b\bar{b}$-initiated production (Figure~\ref{fig:THDMpa_feyn_monojet})
are considered, with the latter only being relevant at large values of $\tan\beta$, where the \monojet\ analysis is not sensitive for the parameter settings
considered in this publication (Section~\ref{sec:summary}).
For larger values of \met and smaller values of $m_a$, the dominant signal process in the $\met+j$ signal regions is the production of two pairs of DM particles ($pp \rightarrow \chichi\chichi$)
via invisible decays of the SM Higgs boson into a pair of mediators that each decay into DM ($h\rightarrow aa \rightarrow \chichi\chichi$).
Depending on the parameter space, the mediators may be real or virtual. The jet arises from the parton shower.
This process is illustrated in Figure~\ref{fig:THDMpa_feyn_haa}.
Additional, though sub-dominant, signal contributions to the \monojet\ signal regions arise from $\met+Z(q\bar{q})$ and $\met+h(\bbbar)$ production
in parameter regions where invisible decays of the SM Higgs boson are kinematically forbidden.
Further, minor contributions arise from the $p p \rightarrow \ttbar+a$ (Figure~\ref{fig:THDMpa_feyn_res_assoc}) and $p p \rightarrow tW + a$ (Figure~\ref{fig:THDMpa_feyn_mettw}) processes.

% End of text imported from the .//sections/input_analyses_metj.tex input file

% The next lines are included from the .//sections/input_analyses_hinv.tex input file
\subsubsection{\hinv}
\label{subsubsec:hinv}

A statistical combination of all ATLAS direct searches for invisible decays of the Higgs boson was published in Ref.~\cite{HIGG-2021-05}.
It is based on five independent searches relying on 139~\ifb of proton--proton collision data collected with the ATLAS detector at a centre-of-mass energy of $\sqrt{s}=13$~\TeV during LHC Run 2.
These searches target Higgs boson production via the vector-boson fusion (VBF), VBF with a photon, gluon-gluon fusion,
associated production with a vector boson, and associated production with \ttbar,
respectively in the $\mathrm{VBF}+\met$~\cite{EXOT-2020-11}, $\mathrm{VBF}+\gamma+\met$~\cite{EXOT-2021-17},
$\met+j$~\cite{EXOT-2018-06}, $\met+Z(\ell\ell)$~\cite{HIGG-2018-26}, and $\met+\ttbar$~\cite{SUSY-2019-12} final states.
The results from the Run 2 searches are further combined statistically with the set of constraints on invisible Higgs decays obtained from searches and measurements
targeting multiple production and decay channels with up to 4.7~\ifb of $pp$ collision data at $\sqrt{s} = 7$~\TeV and 20.3~\ifb at $\sqrt{s} = 8$~\TeV~\cite{HIGG-2015-03},
yielding the most sensitive direct constraint to invisible Higgs boson decays in ATLAS.

Among the direct searches, the VBF production of Higgs bosons decaying into invisible particles using the full Run 2 data sample is the most sensitive one,
setting an observed (expected) upper limit on the invisible branching ratio of 0.145 (0.103) at 95\% confidence level (CL).
Events are selected using \met triggers and the analysis requires $\met > 160$~\GeV and two, three or four jets with $\pT > 25$~\GeV.
The leading and sub-leading jets must have $\pT > 80$~\GeV and 50~\GeV, respectively.
Additional requirements on the angular separation of the two jets are applied to enhance the sensitivity to VBF production.
In particular, the two leading jets are required to be well separated in $\eta$.
Lepton and $b$-jet vetoes are applied to reduce contamination from $W$+jets and top-quark backgrounds, respectively.
Sixteen orthogonal signal regions are defined based on the values of \met, the jet multiplicity
and the two- and three-jet invariant masses in two-jet and three- or four-jet regions, respectively.
The dominant background processes are $Z(\nu\nu)$+jets and $W(\ell\nu)$+jets production,
where in the latter process the charged lepton $\ell$ is not detected or mis-identified.
These backgrounds are evaluated simultaneously using well populated control regions in the one-lepton and two-leptons channels.
Such extrapolation is made possible due to the use of a dedicated theoretical calculation at next-to-leading-order in the relevant phase space~\cite{Lindert:2022ejn}.
The multijet background is directly estimated from data.

An upper limit on the $h\rightarrow$ invisible branching ratio of 0.113 $\left(0.080^{+0.031}_{-0.022}\right)$ is observed (expected) at 95\% CL.
This upper limit is used directly to determine the excluded parameter regions in the 2HDM+$a$ based on the predicted $h\rightarrow \chi\bar{\chi}$ branching ratio
for each point in the benchmark scenarios in Section~\ref{sec:thdma}.
%

% End of text imported from the .//sections/input_analyses_hinv.tex input file

% The next lines are included from the .//sections/input_analyses_metx_36ifb.tex input file
\subsubsection{Additional searches using 36~\ifb~of $\sqrt{s}=13$~\TeV $pp$ collision data}

Results from three searches using 36~\ifb~of $\sqrt{s}=13$~\TeV $pp$ collision data,
which were already included in the summary of 2HDM+$a$ constraints in Ref.~\cite{EXOT-2017-32}, are shown for completeness.
The $\met+Z(q\bar{q})$ search~\cite{EXOT-2016-23} targets final states with $\met > 150$~\GeV and a hadronically decaying $W$ or $Z$ boson candidate.
The vector-boson candidate is reconstructed as a single large-$R$ jet with $\pT > 250$~\GeV in a boosted topology ($\met>250$~\GeV)
or from two small-$R$ jets with $\pT > 20$~\GeV in a resolved topology. In the both cases, a lepton veto is applied.
Several signal regions are defined according to the $b$-jet multiplicity.
The normalisations of the dominant backgrounds from \ttbar and $W/Z+$jets production are constrained using a simultaneous fit to the \met distributions
in the signal and dedicated control regions.

The $\met+\bbar$ search~\cite{SUSY-2016-18} targets events with $\met > 180$~\GeV and at least two $b$-jets.
The azimuthal separations between the $b$-jets and the \met direction are exploited to enhance the separation between the signal
and the irreducible background from $Z(\nu\bar{\nu})+\bbar$ events, which is constrained using data in a dedicated control region.
The results are extracted from a likelihood fit to the angular observable $\cos\theta^*_{\bbar} = |\tanh{\Delta\eta_{\bbar}/2}|$,
which depends on the pseudorapidity differences $\Delta\eta_{\bbar}$ between the two $b$-jets.

Searches targeting events with large \met produced with \ttbar are conducted in different final states classified according to the number leptons.
A search in zero-lepton final states targets events in which the $W$ bosons from both the top quarks decay hadronically~\cite{SUSY-2016-18}.
Events are selected based on the presence of at least four energetic jets, at least two of which are $b$-tagged, and relatively high \met.
Requirements on the invariant mass of reclustered large-$R$ jets are imposed to identify events with a boosted $W$-boson or top-quark decay.
The dominant backgrounds from $Z$+jets, \ttbar, and $\ttbar+Z$ production are constrained in dedicated control regions.
A complementary search in one-lepton final states targets events in which one of the $W$ bosons decays leptonically~\cite{SUSY-2016-16}.
Events are required to contain at least four energetic jets, at least one of which is $b$-tagged, one isolated lepton, and large \met.
They are also required to have at least one hadronic top candidate with invariant mass loosely compatible with the mass of the top quark.
Requirements on the azimuthal angle between the lepton and \ptmiss and on the angular separation $\Delta\phi(\mathrm{jets},\ptmiss)$
are used to suppress the background contamination of the signal regions.
All background processes involving top quarks are estimated in dedicated control regions.

% End of text imported from the .//sections/input_analyses_metx_36ifb.tex input file

\subsection{Searches for visible mediator decays}
%

% The next lines are included from the .//sections/input_analyses_htb.tex input file
\subsubsection{\htb}
\label{subsubsec:htb}

This search targets the production of heavy charged Higgs bosons, $H^{\pm}$, with masses in the range 0.2--2.0~\TeV together with a top and a bottom quark
with the charged Higgs boson decaying into a top and a bottom quark, $pp\rightarrow tbH^{\pm}(tb)$~\cite{HDBS-2018-51}.
Events are pre-selected using single-lepton triggers
and are required to contain exactly one electron or muon with $\pT>27$~\GeV and at least five jets with $\pT>25$~\GeV,
consistent with the semileptonic decay of one of the top quarks.
At least three of the jets are required to be identified as a $b$-jet to suppress the large backgrounds from multijet production.
The selected events are further classified into four separate regions according to the number of reconstructed jets (j) in an event
and number of $b$-jets (b) among them, referred to as 5j3b, 5j$\ge$4b, $\ge$6j3b, and $\ge$6j$\ge$4b.
A neural network is used to enhance the separation between signal and background.
The output distributions of the neural network are used in a fit to extract the amount of \htb signal in data.

The dominant backgrounds for this search are composed of \ttbar+jets events, including \ttlin, \ttbin\ and \ttcin, and single top-quark production in the $Wt$ channel.
Both these processes and smaller background contributions are modelled using MC simulation.
Data-driven corrections obtained in an additional $\ge$5j2b region are applied to the simulation for the leading backgrounds
via a reweighting procedure to improve the modelling of the transverse momentum distributions of additional jet emissions
and of kinematic regions with high jet multiplicities~\cite{ATL-PHYS-PUB-2018-009, TOPQ-2018-18}.
After the reweighting, the final \ttbin\ and \ttcin\ normalisation factors are extracted from the fit to data.

Both the model-independent upper limits on the cross-section times branching ratio for the signal process and the model-dependent exclusion contours on specific benchmarks,
including a type-II 2HDM in the alignment limit without DM, were derived~\cite{HDBS-2018-51}.
These results can be straightforwardly reinterpreted in the context of the 2HDM+$a$ as the dominant production modes
and hence the production cross-sections of the charged Higgs bosons are identical in both of the models.
This was verified by comparing the simulated predictions of the 2HDM and 2HDM+$a$ benchmarks for a range of relevant kinematic variables.
The simulated cross-sections are scaled to their NLO values calculated for the 2HDM+$a$ predictions, in the four-flavour scheme
to be consistent with the modelling choices outlined in Section~\ref{sec:samples}.
These values are on average 20\%--30\% smaller than the corresponding NLO cross-sections calculated in the five-flavour scheme used in Ref.~\cite{HDBS-2018-51},
in accordance with the results in Ref.~\cite{LHCHiggsCrossSectionWorkingGroup}.
The branching ratios of the charged Higgs bosons differ between the \thdma and the 2HDM without DM due to additional possible decay modes of the charged Higgs bosons in the 2HDM+$a$.
Hence the exclusion limits are rescaled by the ratio of branching ratios in the \thdma and the 2HDM, for which the original exclusion limits are derived,
to obtain the exclusion limits for the 2HDM+$a$.

% End of text imported from the .//sections/input_analyses_htb.tex input file

% The next lines are included from the .//sections/input_analyses_tttt.tex input file
\subsubsection{\tttt}

This search specifically targets \ttbar-associated production of heavy scalar or pseudo-scalar Higgs bosons $A/H$ decaying into \ttbar ($\ttbar+A/H\rightarrow \ttbar\ttbar$)~\cite{EXOT-2019-26}.
It is based on and extends the analysis strategy of Ref.~\cite{TOPQ-2018-05} to increase the sensitivity to $A/H$ production.
Single-lepton and the dilepton triggers are used to collect the data on which the search is based.
Events are required to contain either a same-sign lepton pair or at least three leptons ($\ell=e,\mu$).
This includes electron or muons from leptonic $\tau$-lepton decays.
Electrons and muons are required to have $\pT>28$~\GeV.
A baseline signal region is defined by additionally requiring the presence of at least six jets with $\pT>25$~\GeV,
at least two of which must be $b$-tagged,
and $H_T>500$~\GeV, where $H_T$ is defined as the scalar sum of the transverse momenta of all leptons and jets in the event.
A multivariate discriminant based on a BDT is used to separate between SM \tttt~production and background processes,
using event-level information such as jet and $b$-jet multiplicity and additional kinematic variables.
The BSM search relies on a second BDT to distinguish between BSM and SM four-top production.
This second BDT is parameterised as a function of the mass of the heavy Higgs boson by introducing the mass as a labelled input in the training~\cite{Baldi:2016fzo}.

The main, irreducible backgrounds arise from the production of a \ttbar pair together with a boson and additional jets ($\ttbar+W+$jets, $\ttbar+Z+$jets, $\ttbar+h+$jets).
They are estimated by using MC simulations with additional data-driven corrections applied in the case of $\ttbar+W+$jets production.
Smaller, reducible backgrounds arise mostly from $\ttbar+$jets and $tW+$jets production with mis-identified charge, fake and non-prompt leptons.
These smaller backgrounds are estimated from data using dedicated control regions.
%

% End of text imported from the .//sections/input_analyses_tttt.tex input file

% The next lines are included from the .//sections/input_analyses_hlrs.tex input file
\subsubsection{Exotic Higgs boson decays $h\rightarrow aa \rightarrow f\bar{f} f'\bar{f'}$}
\label{subsubsec:hlrs}

Various complementary searches target decays of the $m_h = 125$~\GeV SM Higgs boson to a pair of light pseudo-scalars, which subsequently decay to fermions,
$h\rightarrow aa \rightarrow f\bar{f} f'\bar{f'}$. The searches target final states with different types of fermions and provide sensitivity to different ranges of the pseudo-scalar mass $m_a$.

A search using 139~\ifb of $\sqrt{s}=13$~\TeV $pp$ collision data in the $b\bar{b}\mu^+\mu^-$ final state targets pseudo-scalars in the range of 16~\GeV $\leq m_{a} \leq$ 62~\GeV~\cite{HDBS-2021-03}.
The di-muon invariant mass is the variable of interest in this search, which is probed for a resonant enhancement over the SM expectation.
The dominant backgrounds in the analysis arise from the Drell--Yan di-muon process together with $b$-quarks
and SM \ttbar production where each of the $W$ bosons from the two top quarks decays into a muon and a neutrino.

A search using 36~\ifb of $\sqrt{s}=13$~\TeV $pp$ collision data targeting the $b\bar{b}b\bar{b}$ final state provides sensitivity to pseudo-scalars
in the mass range 20~\GeV $\leq m_{a} \leq$ 60~\GeV~\cite{HIGG-2017-05}.
It targets Higgs boson production in association with a leptonically decaying $W$ (one-lepton channel) or $Z$ boson (two-lepton channel).
Several kinematic variables, including the reconstructed masses in the decay $h \rightarrow aa \rightarrow 4b$, are used as input to a BDT that is trained
to distinguish signal from background events. The dominant background process in the one-lepton signal regions arises from \ttbar production with additional jets,
while the dominant background component in the signal regions with two leptons is due to $Z$+jets production.
The BDT output distribution is used as the observable of interest in the final likelihood fit.
The search is optimised for resolved final states in which the two $b$-jets from each of the $a \rightarrow b\bar{b}$ decays can be reconstructed as two individual
small-$R$ jets. This limits the sensitivity of the search for masses $m_a < 30$~\GeV,
a regime where the two $b$-jet pairs are increasingly likely to be merged into a single large-$R$ jet.
Such merged final states are the target of a complementary search~\cite{HDBS-2018-47} on the same data sample. However, this search provides little additional sensitivity to the \thdma
in comparison to the other searches discussed in this section and is therefore not considered in this publication.

The mass range 3.7~\GeV~$\leq m_{a} \leq$ 50~\GeV~is probed by a search on 20.3~\ifb of $\sqrt{s}=8$~\TeV $pp$ collision data targeting $\mu^+\mu^-\tau^+\tau^-$ final states~\cite{HIGG-2014-02}.
The search probes resonant enhancements in the di-muon invariant mass spectrum.

Finally, masses $m_a \geq 1$~\GeV are probed by two searches targeting final states with four charged leptons ($\ell = e, \mu$)
on 36~\ifb~\cite{EXOT-2016-22} and 139~\ifb~\cite{HDBS-2018-55} of $\sqrt{s}=13$~\TeV $pp$ collision data, respectively,
Each search is based on two orthogonal regions: a low-mass region covering the mass range 1~\GeV $\leq m_{a} \leq$ 15~\GeV, excluding mass ranges around the J/$\psi$ and $\Upsilon$ resonances,
and a high-mass region covering the mass range 15~\GeV $\leq m_{a} \leq$ 60~\GeV.
Only the low-mass region is sensitive to the \thdma and hence considered in this publication.
For this region, only final states with at least four muons ($\mu^+\mu^-\mu^+\mu^-$) are considered due to their greater branching fraction and the selection efficiency for isolated muons being significantly larger
than that for isolated electrons in this mass range.
The searches are therefore referred as $h\rightarrow aa \rightarrow \mu^+\mu^-\mu^+\mu^-$ in the following.
The dominant background processes for these searches arise from $ZZ^{*} \rightarrow \mu^+\mu^-\mu^+\mu^-$ and $h \rightarrow ZZ^{*} \rightarrow \mu^+\mu^-\mu^+\mu^-$ production.
In both the searches, the observable of interest is the average di-muon invariant mass, $<m_{\mu^+\mu^-}> = (m_{12}+m_{34})/2$, where $m_{12}$ and $m_{34}$ are the invariant masses of the two
di-muon pairs that minimise the di-muon pair invariant mass difference $|m_{12} - m_{34}|$.
In the search conducted on the full 139~\ifb\ data sample, the di-muon masses are required to satisfy 1.2~\GeV $\leq m_{12},m_{34}\leq$ 20~\GeV, excluding the mass ranges of 2.0 -- 4.4~\GeV
and 8.0 -- 12.0~\GeV around the J/$\psi$ and $\Upsilon$ resonances, respectively.
Looser requirements are applied in the search conducted on the partial (36~\ifb) data sample, where di-muon invariant masses 0.88~\GeV $\leq m_{12},m_{34} \leq$ 20~\GeV are allowed.
Hence the latter provides sensitivity in the low $m_a$ range where the former is not sensitive.

Model-independent upper limits on the branching ratio of the decay $h\rightarrow aa \rightarrow f\bar{f} f'\bar{f'}$ are obtained for all searches listed above.
This upper limit is used directly to determine the excluded parameter regions in the 2HDM+$a$ based on the predicted $h\rightarrow aa \rightarrow f\bar{f} f'\bar{f'}$ branching ratio
for each point in the benchmark scenarios in Section~\ref{sec:thdma}.
The branching ratio for the \thdma is calculated at NLO with the $\overline{\mathrm{MS}}$ scheme.

% End of text imported from the .//sections/input_analyses_hlrs.tex input file

% End of text imported from the .//sections/input_analyses.tex input file

%
\section{Systematic uncertainties}
\label{sec:systematics}
%

% The next lines are included from the .//sections/systematics.tex input file
Systematic uncertainties for both the background and signal models are considered in each of the analyses presented in Section~\ref{sec:analyses}.
These uncertainties, and the statistical uncertainties, depend on the event selection, the phase space covered by a given analysis, and its background estimation strategy.
The systematic uncertainties include experimental and theoretical uncertainties. Details of the latter can be found in the indvidual analysis publications referred to in the previous section.
Experimental uncertainties may include uncertainties in the absolute jet energy scales and resolutions,
the jet quality requirements, pile-up corrections, $b$-jet identification efficiencies, and the soft contributions to \met.
Uncertainties in lepton identification and reconstruction efficiencies, energy/momentum scale and resolution are included for events with selected or vetoed leptons.
Uncertainties due to the finite size of the background MC samples and others related to the modelling of the background processes are also included in the analyses.
In all analyses, a luminosity uncertainty of 1.7\%~\cite{ATLAS-CONF-2019-021} is applied to backgrounds derived purely from MC simulation.

The signal modelling is subject to some theoretical uncertainties affecting the production cross-section (normalisation) or the signal acceptance.
They include uncertainties related to the PDF, evaluated following the PDF4LHC recommendations~\cite{Butterworth:2015oua},
and uncertainties related to the choice of renormalisation and factorisation scales.
The latter are derived by varying independently such scales by a factor of $2.0$ and $0.5$ relative to the nominal values used for the MC generation.
Additionally, for the \monozll, \monohbb, and \htb analyses, which enter the statistical combination,
uncertainties in the modelling of initial- and final-state radiation and multi-parton interactions are taken into account.
%
%
%

% End of text imported from the .//sections/systematics.tex input file

%
\section{Statistical combination of results}
\label{sec:comb}
%

% The next lines are included from the .//sections/combination.tex input file
A statistical combination of the \monohbb, \monozll, and \htb analyses is performed and described further in this section.
These are generally the most constraining signatures and
cover complementary regions of the \thdmS\ model parameter space.

The statistical combination is facilitated by the input analyses described in Section~\ref{sec:analyses} being statistically independent.
The \monozll\ analysis places a veto on the presence of $b$-jets, whereas signal selections for the \monohbb and \htb analysis require at least two and three $b$-jets, respectively.
Furthermore, the \htb signal region selections require a charged lepton ($e$ or $\mu$), which is vetoed by the \monohbb selections.
Thus no overlap between the three analysis signal selections is expected.
This was validated on the full data luminosity and additionally by applying the different analysis selections to the simulated signal events of the other two analyses.
No overlap between the signal selections was observed in any of these checks.
There was a negligible ($\ll 1$~\%) event overlap observed between the \htb signal selection and a background selection
used by the \monohbb analysis as a leptonic control region, which has no impact on the combination.

\subsection{Statistical analysis}

The combination of the analyses is performed by constructing their combined likelihood and maximising the corresponding profile likelihood ratio~\cite{Cowan:2010js}.
The common fitted parameter of interest, $\mu$, is the signal strength of a given \thdma signal,
defined as the ratio of the observed to the predicted value of the signal cross-section times branching fraction for the specific signal parameter point being tested.
The exclusion of $\mu=1$ in the combined fit means that the data globally across the analyses are incompatible with the predictions for the signal hypothesis under consideration.
Systematic uncertainties are included in the fit as nuisance parameters (NPs), denoted by $\bm{\theta}$,
and are constrained by Gaussian, Poisson or Log-normal probability density functions.
These encode information from {\em auxiliary} measurements and measure the effect of systematic uncertainties.
The fit model also includes normalisation factors, denoted by $\bm{\lambda}$, which are floated in the fit without constraints
to adjust the agreement with data of background components in their corresponding control region(s).

The likelihood used in the combined fit is given by~\cite{HistFactory}:
\begin{equation} \nonumber
\mathcal{L}(\text{data}|\mu, \bm{\lambda}_{\mu}, \bm{\theta}_{\mu}) =
\prod\limits_{c=1}^{N_{\text{cats}}} \mathcal{L}_c(\text{data}|\mu, \bm{\lambda}_{\mu}, \bm{\theta}_{\mu})
\prod\limits_{k=1}^{N_{\text{\text{cons}}}} \mathcal{F}(\tilde{\theta}_{\mu,k}|\theta_{\mu,k})
\end{equation}

where $\bm{\lambda}$ is the vector of normalisation factors,
$\bm{\theta}$ is the vector of nuisance parameters,
$N_{\text{cats}}$ is the number of categories,
$N_{\text{\text{cons}}}$ is the number of constrained NPs,
$\tilde{\theta}_k$ is the global observable corresponding to $\theta_k$,
$c$ is the index for the event categories,
$k$ is the index for the constrained NPs,
and $\mathcal{F}$ denotes a Poisson, a Gaussian or a Log-normal distribution depending on the type of uncertainty.

The 95\% CL limits are obtained using the CLs frequentist formalism~\cite{ReadCLs} with the profile likelihood ratio test statistic ($q_{\mu}$)
implemented using RooStats~\cite{Moneta:2010pm} and RooFit~\cite{RooFit}, defined as~\cite{Cowan:2010js}:

\begin{equation} \nonumber
q_{\mu} = \frac{\mathcal{L}(\mu, \hat{\hat{\bm{\lambda}}}_{\mu}, \hat{\hat{\bm{\theta}}}_{\mu} ) }{ \mathcal{L}(\hat{\mu},\hat{\bm{\lambda}}_{\hat{\mu}}, \hat{\bm{\theta}}_{\hat{\mu}} )  },
\end{equation}

where the numerator indicates the values of $\bm{\lambda}_{\mu}$ and $\bm{\theta}_{\mu}$ that maximise $\mathcal{L}$ for a given value of $\mu$,
and the denominator is evaluated for the values $\hat{\mu}$, $\hat{\bm{\lambda}}_{\hat{\mu}}$, $\hat{\bm{\theta}}_{\hat{\mu}}$ which jointly maximise the likelihood.

\subsection{Treatment of uncertainties and their correlations}

There are many sources of uncertainty present in the \monohbb, \monozll, and \htb\ analyses; their correlations are treated as follows.
Most experimental uncertainties, such as those related to the reconstruction of the physics objects are correlated across search channels,
as are the uncertainties in the integrated luminosity and the modelling of pile-up. This includes the uncertainties from electrons, muons, \met, and the jet energy response.
The assessment of the correlations of uncertainties stemming from $b$-jet identification are complicated by differing choices of algorithm and operating point,
hence these are not correlated.
Finally, a handful of experimental systematic uncertainties that are moderately constrained in a particular analysis are not correlated
to avoid introducing any phase-space-specific biases.
Different assumptions on the correlation of the uncertainties related to jet, \met, and $b$-jet indentification and the moderately constrained uncertainties
were tested separately to assess their impact on the observed exclusions. The effect on the observed exclusions was found to be negligible.

Uncertainties are assessed on the signal simulation and background modelling for each of the analyses to be combined.
Dedicated signal simulations are performed for each of the final states, as they often probe very different kinematic regions of phase space.\footnote{Each analysis is assessed
at each model parameter point, unless the sensitivity of one is known to be negligible in that region of parameter space, in which case it may be omitted and its sensitivity set to zero.}
The resulting theoretical uncertainties are found to be small and often completely negligible, and are considered to be uncorrelated.
The uncertainties related to the estimate of the backgrounds are considered to be uncorrelated amongst the analyses.
This is motivated by their different sources of leading background, the different kinematic phase space probed,
and the wide-spread use of data-derived and analysis-specific methods of background estimation.

\subsection{Impact of uncertainties}
%

% The next lines are included from the .//sections/errorsplit_800_450.tex input file
Inevitably, the contributions of the many uncertainties on the analysis combination vary across the model parameter values, due to the differing signal kinematics and the varying sensitivities of the individual analyses.
The contributions to the total uncertainty in the best-fit signal strength from the statistical and systematic uncertanties
are shown in Table~\ref{tab:sys:BR} for \ma = 450~\GeV, \mH = 800~\GeV, \tanb = 1.0 and $\sin\theta = 0.35$.
This particular signal is narrowly excluded by the combination but is not by any single input analysis, and all three analyses contribute some sensitivity.
For this signal, the statistical uncertainty is comparable to (but slightly smaller than) the systematic component, which is broken down into three categories of sources: theoretical, experimental and MC statistical uncertainties. For each category, the uncertainty is assessed by fixing the corresponding uncertainties in a fit and subtracting the resultant uncertainty from the total in quadrature. The theoretical uncertainties, which stem predominantly from uncertainties in the background modelling, are slightly smaller than the experimental ones. The experimental uncertainty is further subdivided into those on each of the reconstructed physics objects, amongst which the largest contributions come from jet and \met\ uncertainties.

The most important uncertainties follow directly from those of the individual input analyses. For the background modelling uncertainties,
the largest components are $ZZ$ modelling from the \monozll analysis, \ttbar uncertainties affecting \monohbb,
and uncertainties from the production of \ttbar with additional $b$-quarks, which impact the \htb search.
Among the experimental systematic uncertainties, the largest sources are lepton systematic uncertainties impacting \monozll,
uncertainties related to jets and \met affecting \monohbb,
and systematic uncertainties related to the identification of $b$-jets in the case of the \htb analysis.

\begin{table}[htbp]
\centering
\caption{\label{tab:sys:BR} Summary of the uncertainties $\Delta\mu$ in the best-fit signal strength on a signal ($m_A=800$~\GeV, $m_a=450$~\GeV, $\tanb = 1$, $\sint = 0.35$),
obtained by fixing the corresponding nuisance parameters to their best-fit values,
and subtracting the square of the resulting uncertainty from the square of the total uncertainty to evaluate $(\Delta\mu)^2$.
The statistical uncertainty component is obtained by fixing all nuisance parameters except free-floating background normalisation factors to their best-fit values, and quantitifes the impact of the limited data yields in the signal and control regions.
Note the total uncertainty does not equal the sum of the individual contributions added in quadrature due to correlations between the systematic uncertainties.}

\renewcommand{\arraystretch}{1.3}
\begin{tabular}{lr}
\hline\hline
Uncertainty source                               & $\Delta\mu \cdot 100$  \\
\hline
Statistical uncertainty                          & 25.0 \\
\hline
Systematic uncertainties                         & 27.6 \\
~~~~~Theory uncertainties                        & 16.2 \\
~~~~~~~~~~Signal modelling                       & 2.8 \\
~~~~~~~~~~Background modelling                   & 15.9 \\
~~~~~Experimental uncertainties (excl.\ MC stat.) & 18.8 \\
~~~~~~~~~~Luminosity, pile-up                     & 3.9 \\
~~~~~~~~~~Jets, \met                             & 12.3 \\
~~~~~~~~~~Identification of $b$-jets             & 9.1 \\
~~~~~~~~~~Electrons, muons                       & 6.1 \\
~~~~~MC statistical uncertainty                  & 9.3 \\
\hline
Total uncertainty                                & 37.2 \\
\hline\hline
\end{tabular}
\end{table}

% End of text imported from the .//sections/errorsplit_800_450.tex input file

% End of text imported from the .//sections/combination.tex input file

%
\section{Summary of constraints on the 2HDM+$a$}
\label{sec:summary}
%

% The next lines are included from the .//sections/results.tex input file
\subsection{Scenario 1: $\ma-\mA$ planes}
\label{sec:thdma_sum_mAma}

The exclusion contours for the $m_A-m_a$ scans with \sint$=0.35$ and \sint$=0.7$,
which correspond to Scenarios 1a and 1b in Section~\ref{sec:thdma}, respectively,
are shown in Figure~\ref{fig:scan_mA_ma}.
In the upper sub-figures, the exclusion regions for the statistical combination of the \monohbb, \monozll, and \htb searches are shown,
along with the exclusion regions from the three individual searches entering the combination.
The \monozll\ and \monohbb\ searches dominate the sensitivity across a large fraction of the two parameter planes,
which is largely due to the resonant production of the (pseudo-)scalars according to the diagram in Figure~\ref{fig:THDMpa_feyn_gg_res}.
Their sensitivities depend on both the pseudo-scalar Higgs boson and mediator masses.
For \sint$=0.35$ (Scenario 1a), the maximum reach obtained for $m_a$ is up to 560~\GeV, if the $A$ boson mass is set to 1.2~\TeV,
while for $m_a=150$~\GeV values of \mA between 250~\GeV and 1.55~\TeV are excluded.
For both the \sint choices, in the lower left area, the \monozll\ limit reaches closer to the $m_A = m_a$ line than the \monohbb\ limit.
This is because \monozll\ can probe lower \met\ values, whereas \monohbb\ requires a higher \met\ threshold due to the use of a \met\ trigger
and due to the mass difference between the $Z$ and Higgs bosons.
For both \sint choices, but most notably for \sint$=0.7$ (Scenario 1b), an increase in the exclusion power of the $\met+h$ searches is observed at larger values of $m_A$ and low values of $m_a$.
This is due to an increase of the cross-section of the non-resonant $a^*\rightarrow ah$ process, without resonant $A$ production.
There is no equivalent process for the $\met+Z$ signature.

The \htb search provides complementary sensitivity to the \monozll\ and \monohbb\ searches, excluding the range $m_A \lesssim 700$~\GeV for \sint$=0.35$.
Its exclusion contour shows only a moderate dependence on $m_a$ as this search does not probe the production of the pseudo-scalar mediator directly
and is therefore only indirectly affected by the choice of $m_a$ via its effect on the relative branching ratio to $tb$
compared with the branching ratios for other possible decay modes, such as $H^{\pm}\rightarrow aW^{\pm}$.
The effect of this reduction of the branching ratio to $tb$ is visible in Scenario 1b,
where the limits from the \htb search are slightly weaker at low values of \ma where, for example, the decay into $aW^{\pm}$ is kinematically allowed.
The combination of the \htb with the \monozll\ and \monohbb\ searches increases the excluded parameter space, especially the excluded \ma range for $m_A\approx 800$~\GeV (Scenario 1a)
and for $m_A\approx 700$~\GeV (Scenario 1b).

The exclusion regions from other searches not entering the combination are added in the lower sub-figures of Figure~\ref{fig:scan_mA_ma}.
The limit on the branching ratio for invisible Higgs boson decays constrains very low values of $m_a$,
as searches for invisible Higgs boson decays are only sensitive to light $a$ bosons decaying into invisible particles.
The \monohgamgam search probes a similarly shaped, albeit smaller, region in parameter space compared with the \monohbb search due to the smaller branching ratio
to $\gamma\gamma$ compared with \bbbar.
However, its sensitivity exceeds that of the \monohbb search for low values of \mA because it does not rely on \met\ triggers exclusively and hence is able to probe smaller values of \met.
Like in the case of the \monohbb search, a significant increase in sensitivity for high values of \mA is found due an increase of the cross-section of the $a\rightarrow ah$ process.
The \monohtautau search is only interpreted in the scenario with \sint$=0.35$. Its exclusion contour shows a similar $m_A-m_a$ dependence as that of the \monohbb search
but its sensitivity is notably lower due to smaller Higgs boson branching ratio to $\tau\tau$ compared with \bbbar final states.

The exclusion contour for the $\met+tW$ search shows a shape similar to those of the \monozll\ search for both \sint choices, although the overall exclusion region is smaller.
Its observed exclusion is weaker than the expected sensitivity due to a small (less than $2\sigma$) excess in the 2-lepton channel~\cite{EXOT-2018-43}.
The sensitivity of the $\met+tW$ search is greater for larger values of \sint~\cite{Pani:2017qyd}.

The sensitivity of the $\met+j$ search in the $\ma-\mA$ plane is notably different from those of the $\met+Z$ and $\met+h$ searches
due to the absence of resonant production diagrams for this signature.
The signal cross-section for this signature, and hence its sensitivity to the \thdma is affected by interference effects
between the non-resonant contributions involving the two pseudo-scalars $a$ and $A$.
The impact of the interference depends notably on the values of both the \ma and \mA and is more pronounced for higher values of \sint due to the larger $a-A$ mixing~\cite{Bauer:2017ota}.
In particular for signal hypotheses characterised by a small mass difference between $A$ and $a$ ($m_A \approx m_a$), the interference is destructive, leading to smaller signal cross-sections
and hence a reduced sensitivity of the $\met+j$ signature to the 2HDM+$a$.
This effect is visible for both $\sint=0.35$ and $\sint=0.7$.
For the scenario with $\sint=0.35$, the $\met+j$ search excludes values of \ma up to 600~\GeV for $m_A \approx 200$~\GeV and values of \mA up to 800~\GeV for $m_a \approx 100$~\GeV.
For the scenario with $\sint=0.7$, the larger $a-A$ mixing results in higher cross-sections for signal hypotheses with $m_A > m_a$.
For $m_A \approx 1300$~\GeV, the exclusion power in terms of \ma of the $\met+j$ search is comparable with that of the \monozll and \monohbb searches.
A small region for $m_A < m_a$ is also excluded by the $\met+j$ search in the high-$\sint$ scenario.

The \tttt search, like the \htb search, provides complementary sensitivity to the $\met+X$ searches,
with the main difference that it is only sensitive to the \thdma if either \mA or \ma is above the \ttbar production threshold ($m_{A/a} \geq 2 m_t$).
For \sint$=0.35$, the \tttt contour shows a behaviour similar to that of the \htb search. The sensitivity is mostly driven by resonant $A/H$ production and largely independent of $m_a$.
For \sint$=0.7$, the sensitivity of the \tttt search is smaller for $m_a < 2 m_t$ compared with the scenario with $\sint=0.35$ due to the larger $a-A$ mixing
and the fact that in this regime the decay into $a\rightarrow \ttbar$ is kinematically inaccessible.

For completeness, the observed and expected exclusion contours from a $\met+Z(q\bar{q})$ search on 36~\ifb are shown for the scenario with \sint$=0.35$.
This result was already included in Ref.~\cite{EXOT-2017-32}.
The sensitivity of this search is considerably smaller than that of the \monozll\ search due to the larger backgrounds from multijet production in the hadronic decay channel
and due to the smaller data sample on which the search is based.

\begin{figure}[htbp!]
\centering
\subfloat[]{\includegraphics[width=0.49\linewidth]{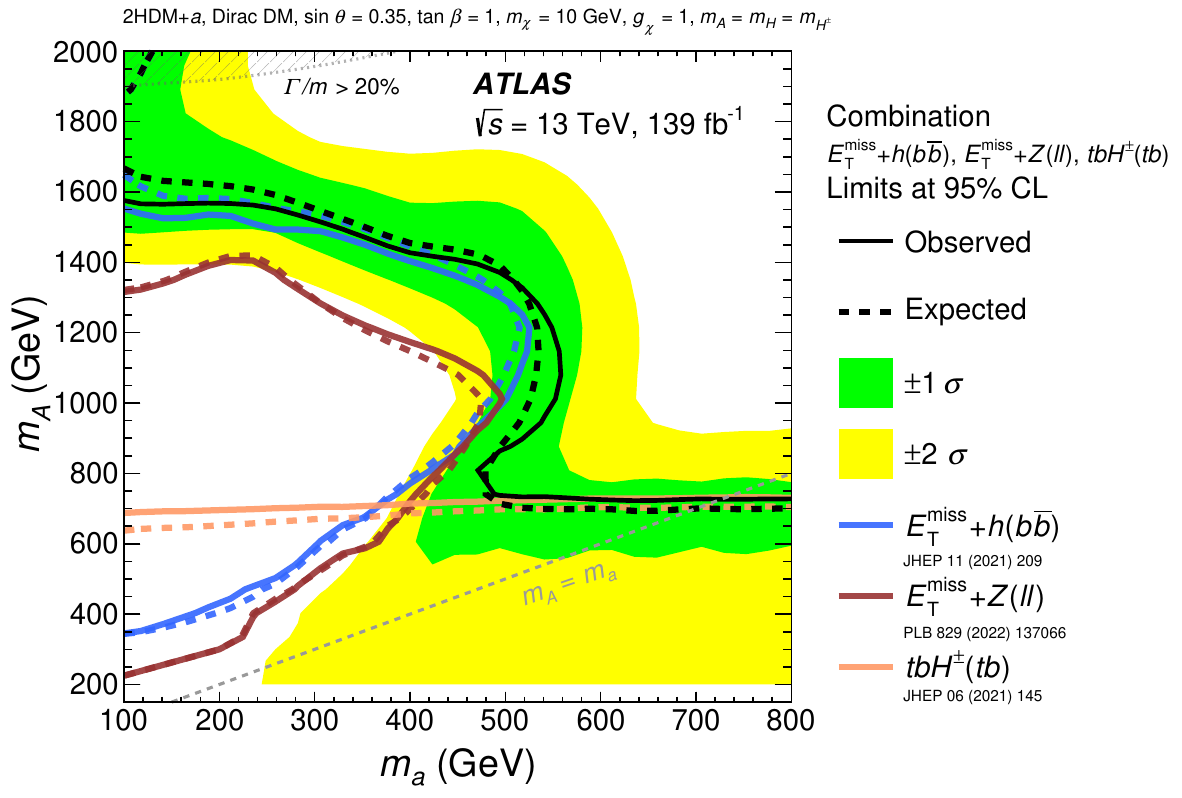}}
\subfloat[]{\includegraphics[width=0.49\linewidth]{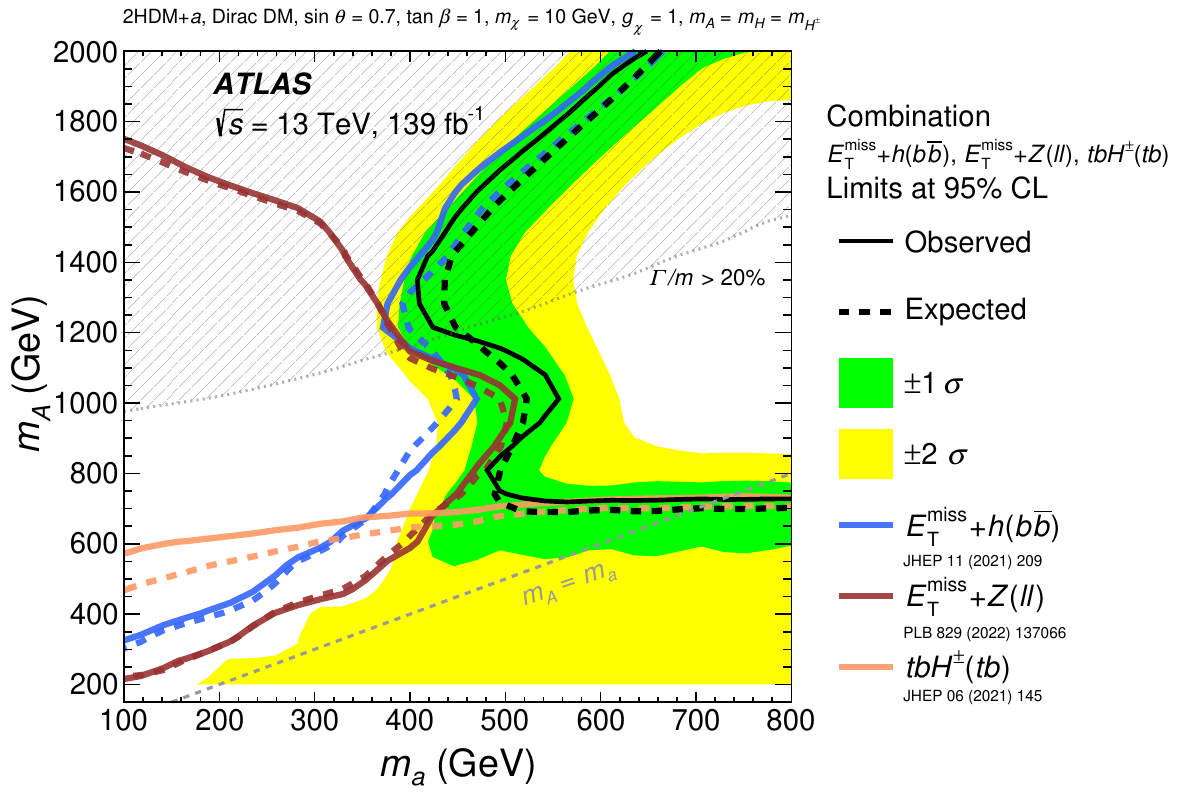}}\\
\subfloat[]{\includegraphics[width=0.49\linewidth]{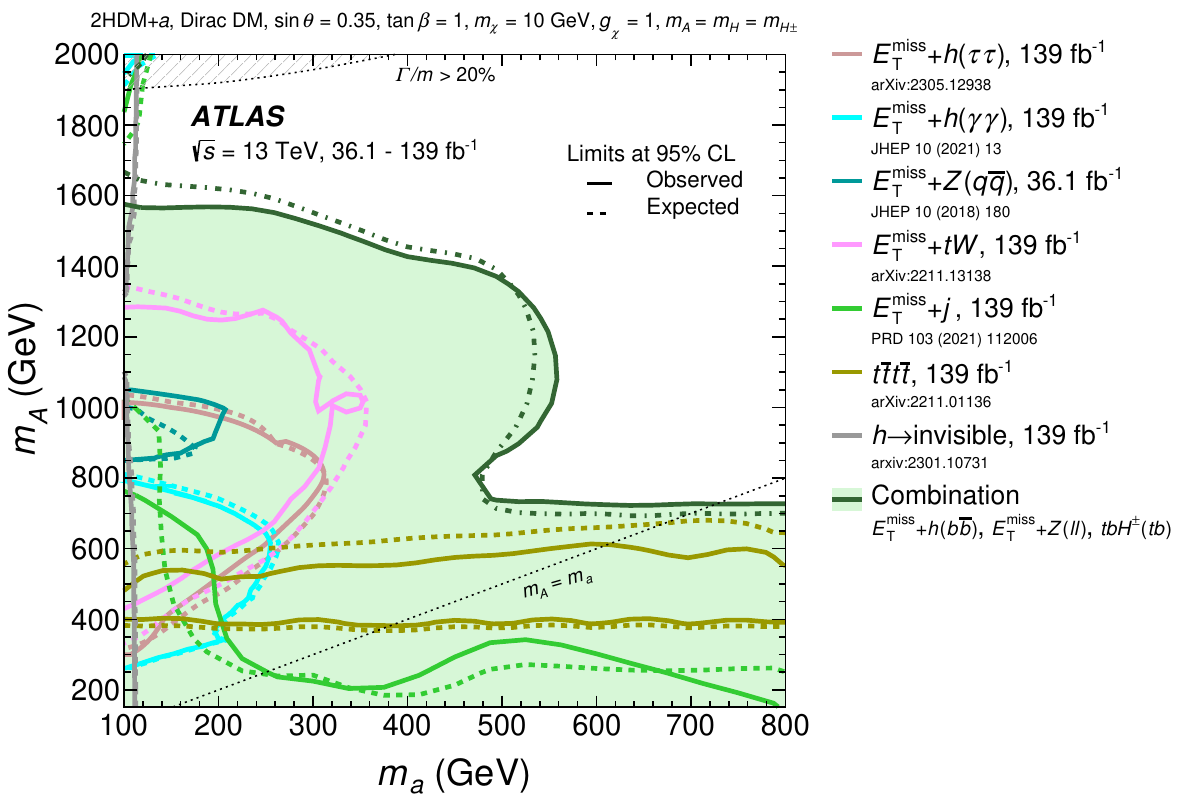}}
\subfloat[]{\includegraphics[width=0.49\linewidth]{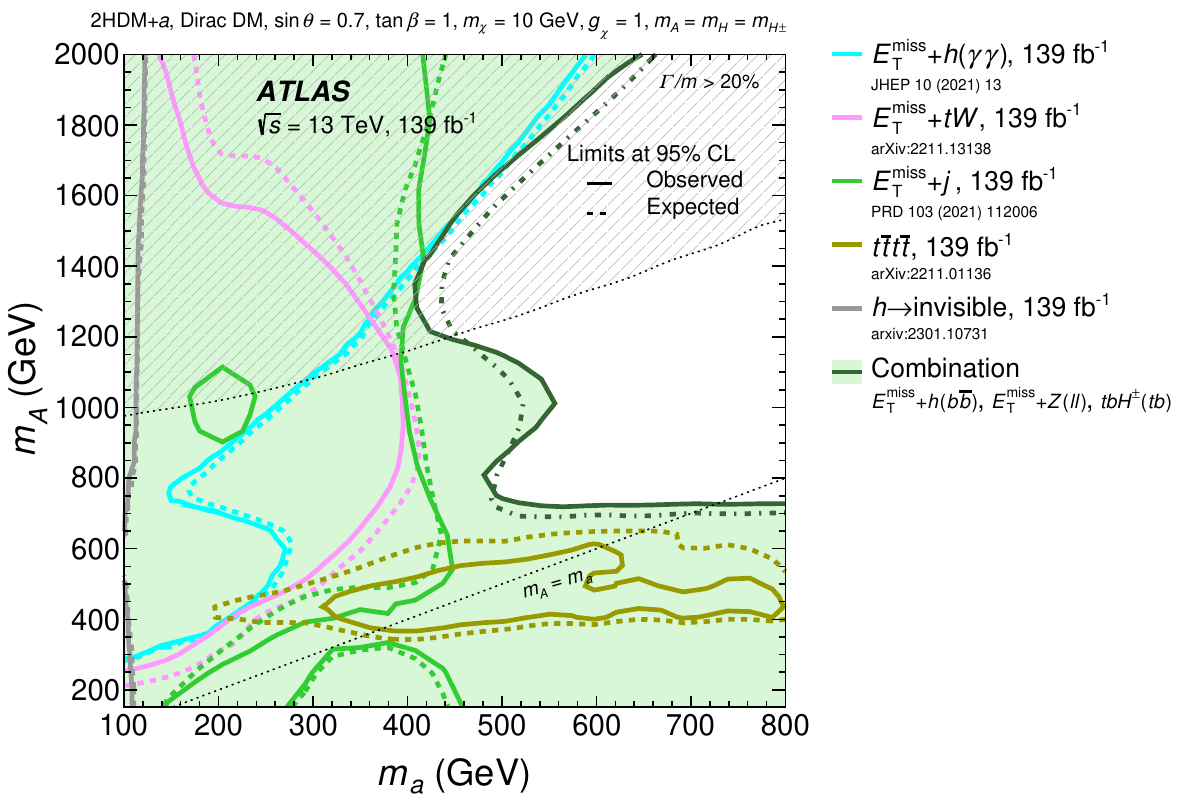}}
\caption{
Observed (solid lines) and expected (dashed lines) exclusion regions at 95\% CL in the $(m_a, m_A)$ plane assuming
(a, c) $\sint=0.35$ (Scenario 1a) and (b, d) $\sint=0.7$ (Scenario 1b).
In the upper sub-figures, the observed (solid lines) and expected (dashed lines) exclusion regions for the statistical combination of the \monohbb, \monozll, and \htb searches
are shown, along with the observed and expected exclusion regions for the three individual searches entering the combination.
The surrounding shaded bands correspond to the $\pm 1$ and $\pm 2$ standard deviation ($\pm 1 \sigma$, $\pm 2 \sigma$) uncertainty in the expected limit of the combined result.
In the lower sub-figures, the results are shown for the combination of the \monohbb, \monozll, and \htb searches (filled area) and additional individual searches.
The individual results from the \monohbb, \monozll, and \htb searches are not shown in this case.
In all four sub-figures, dashed grey regions indicate the region where the width of any of the Higgs bosons exceeds $20\%$ of its mass.
}
\label{fig:scan_mA_ma}
\end{figure}

\FloatBarrier

\subsection{Scenario 2: $\mA-\tanb$ planes}

Exclusion limits as a function of the mass of the pseudo-scalar $A$ and $\tan\beta$ (Scenario 2) are summarised in Figure~\ref{fig:scan_mA_tanB},
again for the two scenarios with \sint$=0.35$ and \sint$=0.7$.
For both the scenarios, a large fraction of the $m_A-\tanb$ plane is excluded by the \monozll search alone.
For higher values of $m_A$, the \monohbb search, and hence the combination of the \monozll, \monohbb, and \htb searches provides the strongest constraints.
The sensitivity of the \monozll and \monohbb searches as a function of \tanb is driven by the transition from $gg$- to $b\bar{b}$-initiated production
with a minimum in sensitivity in the transition region around $\tan\beta \approx 5$.

The $\met+tW$ search probes values of $\tan\beta$ up to 1.5 (\sint$=0.35$) and 2 (\sint$=0.7$).
Its observed exclusion is weaker than the expected sensitivity due to a small (less than $2\sigma$) excess in the two-lepton channel~\cite{EXOT-2018-43}.
Again, the sensitivity of the search is larger for the scenario with $\sint=0.7$ compared with that with $\sint=0.35$~\cite{Pani:2017qyd}.
The \monohtautau search has only been interpreted for the scenario with $\sint=0.7$. Its sensitivity is notably smaller compared with that of the \monohbb search
due to the smaller Higgs boson branching ratio to $\tau\tau$ compared with $b\bar{b}$.
No exclusion contours are shown for the $\met+h(\gamma\gamma)$ search.

The sensitivity of the \htb\ and \tttt searches is largest at low values of \mA and \tanb.
This is due to the larger production cross-section for smaller resonance masses and the preference of third-generation couplings at low values of \tanb.

\begin{figure}[htbp!]
\centering
\subfloat[]{\includegraphics[width=0.49\linewidth]{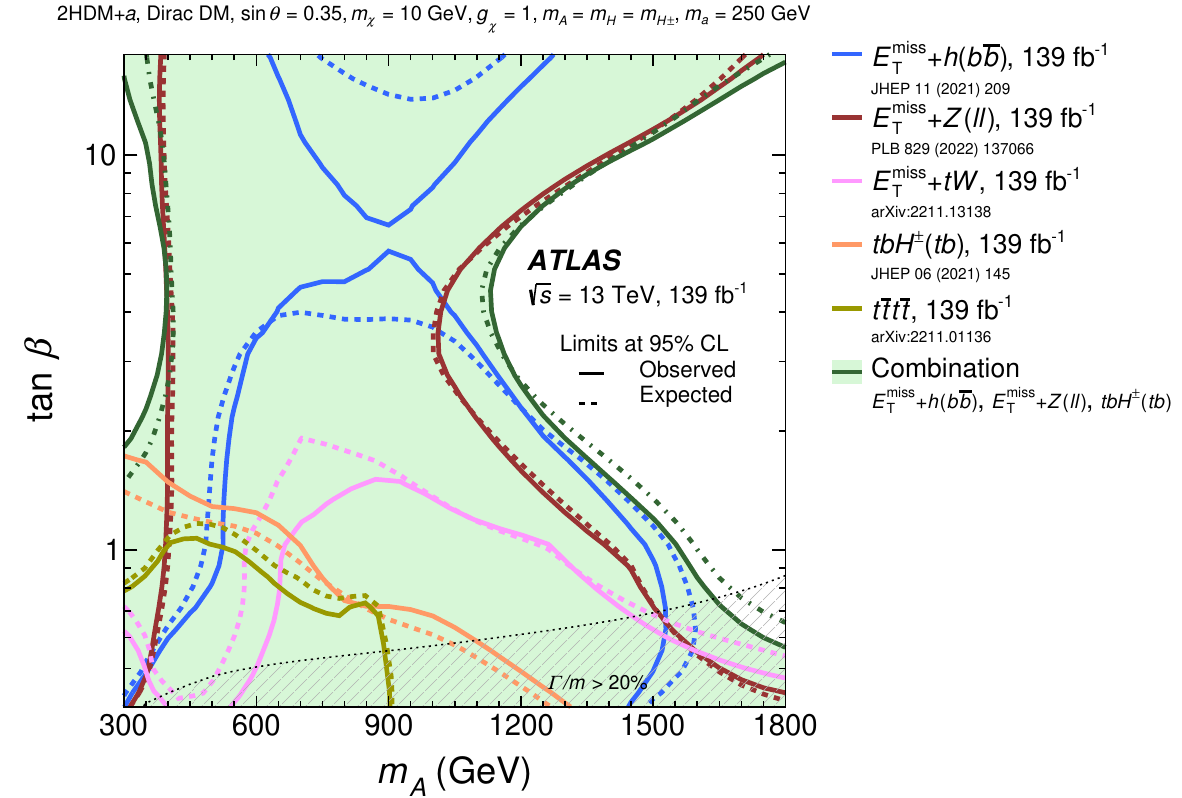}}
\subfloat[]{\includegraphics[width=0.49\linewidth]{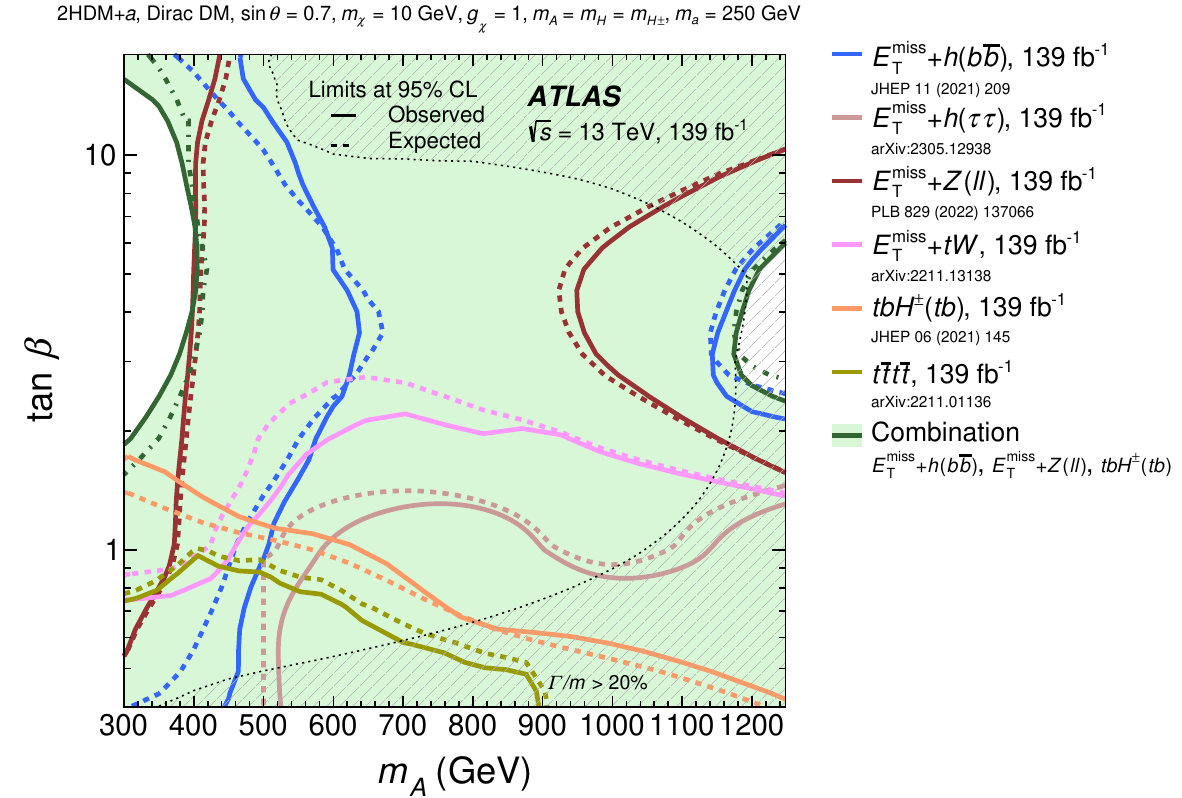}}
\caption{
Observed (solid lines and filled area) and expected (dashed lines) exclusion regions at 95\% CL in the $(\mA, \tan\beta)$ plane assuming
(a) $\sint=0.35$ (Scenario 2a) and (b) $\sint=0.7$ (Scenario 2b).
The results are shown for several individual searches and the combination of the \monohbb, \monozll, and \htb searches.
The dashed grey regions indicate the region where the width of any of the Higgs bosons exceeds $20\%$ of its mass.
}
\label{fig:scan_mA_tanB}
\end{figure}

\FloatBarrier

\subsection{Scenario 3: $\ma-\tanb$ planes}

In Figure~\ref{fig:scan_ma_tanB}, a similar benchmark scenario to that in Figure~\ref{fig:scan_mA_tanB} is shown
with the difference that the mass of the pseudo-scalar mediator $m_a$ rather than $m_A$ is varied (Scenario 3).
Again, the exclusion contours are shown for both the $\sint=0.35$ (Scenario 3a) and $\sint=0.7$ (Scenario 3b).
The strongest exclusion is provided by the \monozll\ search. Its exclusion varies between $m_a\approx 350$~\GeV at $\tan\beta=0.4$ to above 400~\GeV for $\tan\beta \approx 10$.
The \monohbb and \monohgamgam searches exclude a similar \tanb range as the \monozll search but the sensitivity does not reach as high in \ma as that of the \monozll search.
In both the cases, as seen in the $\tan\beta-m_A$ scan for the \monohbb search, a decrease in sensitivity is observed for $\tan\beta\approx 5$
due to the transition from $gg$- to $b\bar{b}$-initiated production.

The $\met+tW$ search probes the range of low \ma and low \tanb values with the sensitivity being slightly higher for the scenario with $\sin\theta=0.7$.
The $\met+j$ search also excludes signal hypotheses characterised by low values of \ma and \tanb.
The sensitivity of this search is higher for the scenario with $\sin\theta=0.7$ compared with that with $\sin\theta=0.35$ due to the higher $a-A$ mixing,
which leads to larger signal cross-sections for $m_A > m_a$, as pointed out in Section~\ref{sec:thdma_sum_mAma}.
Similarly to the $\tan\beta-m_A$ scan, the branching ratio limit on invisible Higgs boson decay provides constraints at very low values of $m_a$, independent of the value of $\tan\beta$.

The sensitivity of the \htb and \tttt searches is complementary to that of the $\met+X$ searches.
It is mostly limited to the low \tanb region and shows only a moderate dependence on $m_a$.

\begin{figure}[htbp!]
\centering
\subfloat[]{\includegraphics[width=0.49\linewidth]{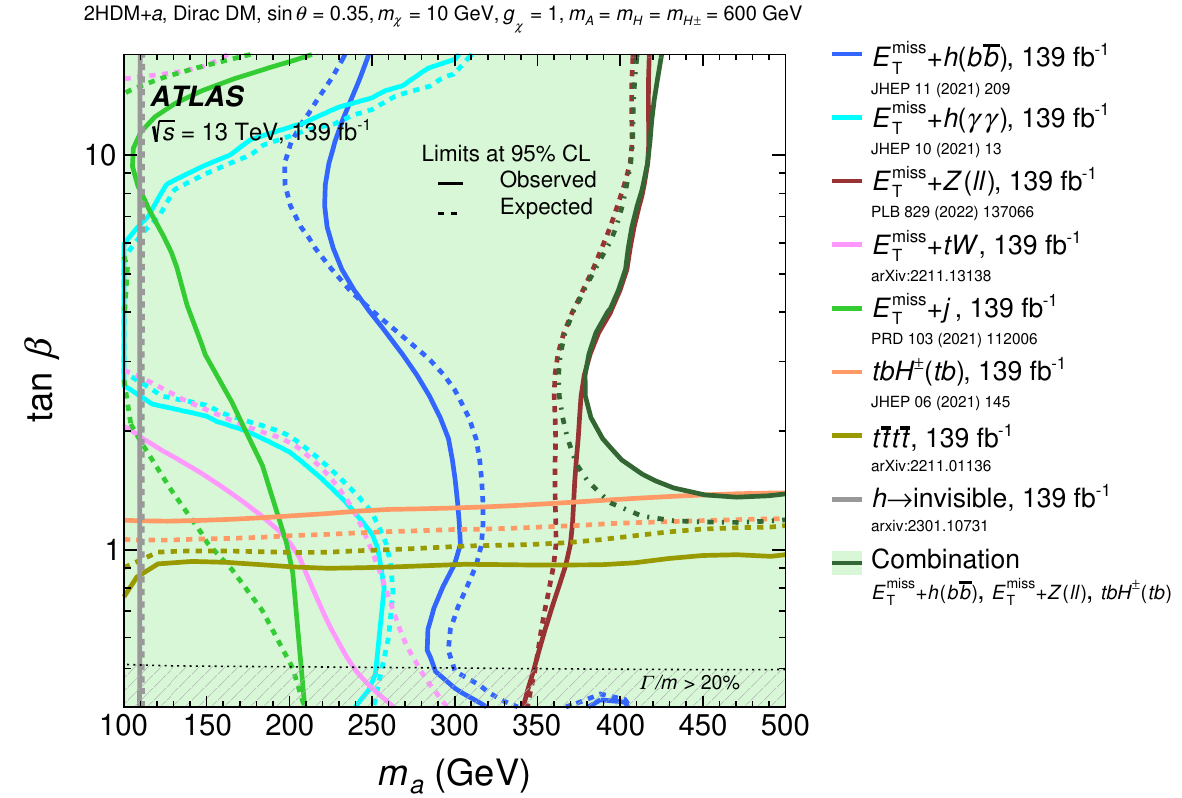}}
\subfloat[]{\includegraphics[width=0.49\linewidth]{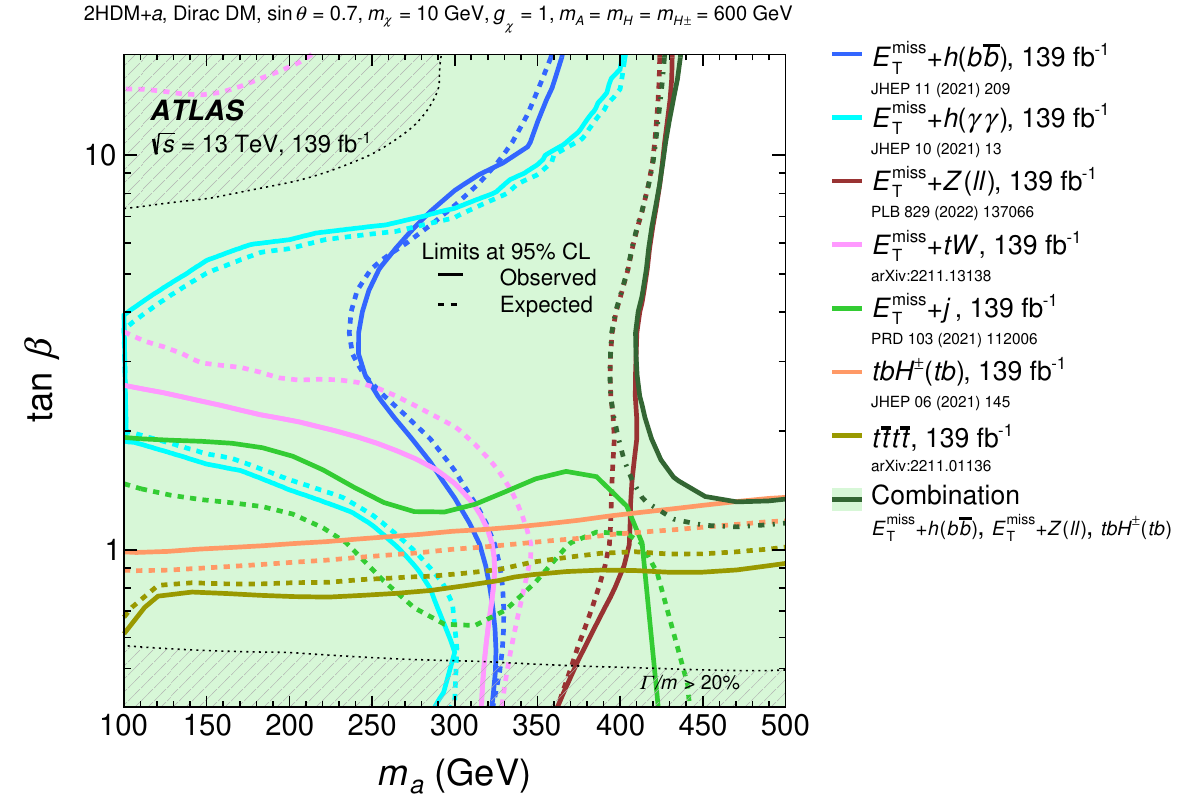}}
\caption{
Observed (solid lines and filled area) and expected (dashed lines) exclusion regions at 95\% CL
in the $(m_a, \tan\beta)$ plane assuming (a) $\sint=0.35$ (Scenario 3a) and (b) $\sint=0.7$ (Scenario 3b).
The results are shown for several individual searches and the combination of the \monohbb, \monozll, and \htb searches.
The dashed grey regions indicate the region where the width of any of the Higgs bosons exceeds $20\%$ of its mass.
}
\label{fig:scan_ma_tanB}
\end{figure}

\FloatBarrier

\subsection{Scenario 4: variation of \sint}

Exclusion limits as a function of \sint for the \hdma\ for the low-mass and high-mass mediator hypothesis,
respectively, are shown in Figure~\ref{fig:scan_sint}.
In the upper row of this figure, results interpreted for the baseline parameter choice $\tan\beta=1$ of Scenario 4 are summarised,
while in the lower row, additional results derived for the alternative choices $\tan\beta=0.5$ or $\tan\beta=50$ are shown.
The sub-figures on the left correspond to the low-mass hypothesis ($m_A = 0.6$~\TeV, $m_a = 200$~\GeV, Scenario 4a),
while those on the right are derived assuming the high-mass hypothesis ($m_A = 1.0$~\TeV, $m_a = 350$~\GeV, Scenario 4b).
The limits are expressed in terms of the ratio of the excluded cross-section to the nominal cross-section of the model.

For the low-mass hypothesis with $\tan\beta=1.0$, the strongest limits in the medium and high \sint range are provided by the \monozll\ and \monohbb\ searches.
The sensitivity of the former monotonically increases as a function of \sint, as the cross-section of the non-resonant and resonant production diagrams,
in Figures~\ref{fig:THDMpa_feyn_gg_res} and~\ref{fig:THDMpa_feyn_gg_box}, respectively, increases with $\sin\theta$.
The same production diagrams for the $\met+h$ signature have very different \sint dependence, as described in Refs.~\cite{Bauer:2017ota,EXOT-2017-32}.
The relative contributions of each diagram are additionally affected by the different \monohbb and \monohgamgam analysis selections.
Both the analyses show a maximum of sensitivity around $\sin\theta\approx 0.5$.
The sensitivities of the $\met+j$ and $\met+tW$ searches, like that of the \monozll search, increase monotonically with \sint
but remain about an order of magnitude below that of the \monozll search across the full \sint range due to the overall lower cross-sections for these processes.
The \htb\ and \tttt signatures show a different \sint dependence compared with the other signatures as they are not directly sensitive to the neutral boson production.
They are particularly sensitive at very small mixing angles, with the \htb sensitivity exceeding those of the \monohbb and \monozll searches, respectively,
for $\sint \lesssim 0.2$.
The results from the $\met+V(q\bar{q})$ search from Ref.~\cite{EXOT-2017-32} are shown for completeness.

For the high-mass hypothesis with $\tanb=1.0$, the mass of the light pseudo-scalar is high enough that the decay $a\rightarrow t\bar{t}$
is kinematically allowed, which introduces an additional \sint dependence to the $\met+Z$ and $\met+h$ analyses interpreted in this scenario.
For this reason, the highest sensitivity for the $\met+Z$ and $\met+h$ analyses is found to be around (or slightly below) the maximal mixing condition ($\theta=\pi/4$).
However, the $\met+h$ signatures have a complex \sint dependence due to the different contributions of resonant and non-resonant processes to the final selection in the two analyses.
The sensitivity of the \monohbb search shows a broad maximum for \sint values below the maximal mixing condition ($\theta = \pi/4$).
The \monohgamgam search instead shows a local sensitivity minimum around $\sin\theta \simeq 0.6$.
The $\sin\theta$ dependence of the $\met+tW$ search is similar to that of the \monohbb and \monozll searches buts its sensitivity is roughly an order of magnitude
below that achieved with the combination of the \monohbb, \monozll, and \htb searches.
In constrast, the $\met+j$ search shows a monotonic increase in sensitivity with increasing $\sin\theta$
and has a similar sensitivity to the \monohbb and \monozll searches for large values of $\sin\theta$.
The \htb signature, similarly to the low-mass mediator hypothesis, shows a constant sensitivity as a function of $\sin\theta$.
Again, the results from the $\met+V(q\bar{q})$ search from Ref.~\cite{EXOT-2017-32} are shown for completeness.

Alternative choices of $\tan\beta=0.5$ or $\tan\beta=50$ for Scenario 4 are explored to highlight the strong \tanb dependence of the exclusion power of searches
with a strong dependence on the Yukawa couplings of the neutral Higgs bosons and the mediator to fermions in a type-II 2HDM.
At low values of $\tan\beta$, the scalars and pseudo-scalars couple preferentially to top quarks, while at high values of $\tan\beta$, couplings to bottom quarks are preferred.
Hence, the results of the \tttt search are shown for $\tan\beta=0.5$.
The sensitivity of the \tttt search is higher for the low-mass compared with the high-mass scenario primarily due to the lower cross-sections for $A/H$ production at higher values of $m_{A/H}$.
In the high-mass scenario, an increase in the \tttt sensitivity is observed for $\sin\theta > 0.5$ due to the increased $a-A$ mixing
and the fact that the mediator mass in this scenario is large enough to allow mediator decays into \ttbar and at the same time considerably below the $A/H$ masses,
which results in the \tttt signal cross-section being completely dominated by $\ttbar + a(\ttbar)$ production.
For completeness, the results from the $\met+t\bar{t}$ and $\met+b\bar{b}$ searches included in Ref.~\cite{EXOT-2017-32} are shown for values of $\tan\beta=0.5$ and $\tan\beta=50$, respectively.

\begin{figure}[htbp!]
\centering
\subfloat[]{\includegraphics[width=0.49\linewidth]{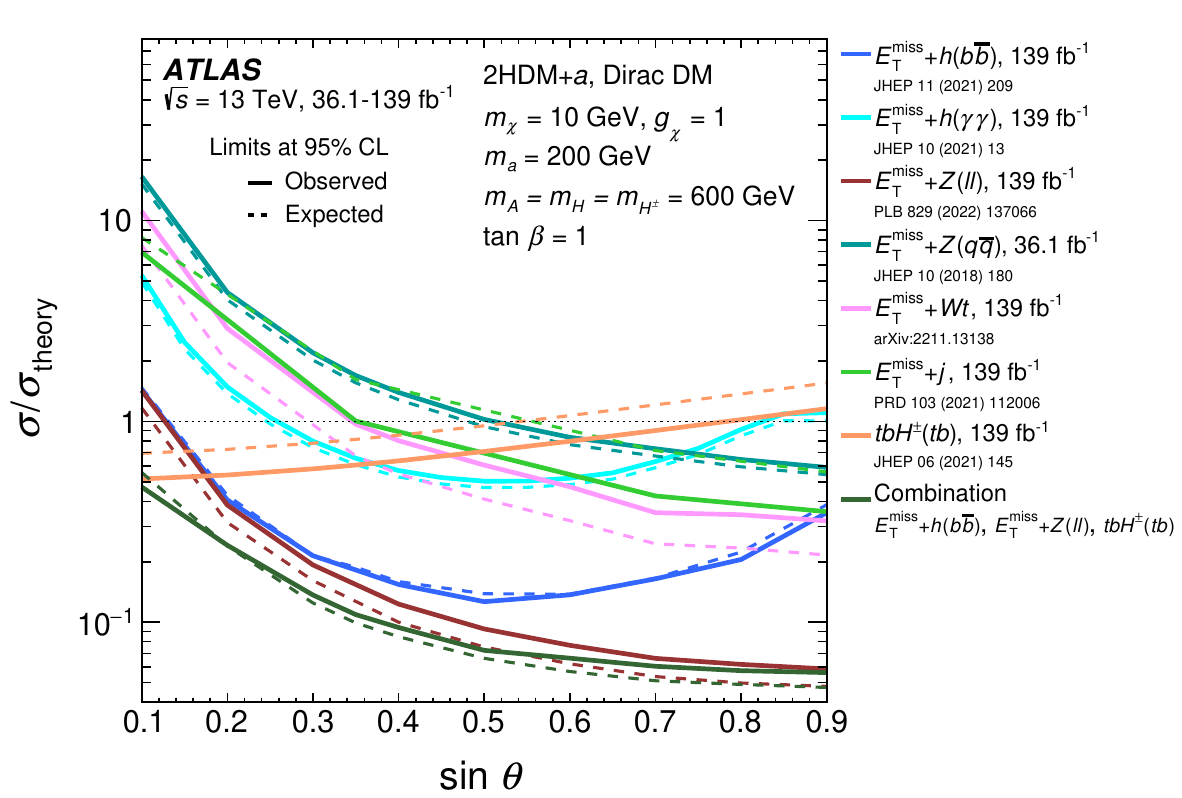}}
\subfloat[]{\includegraphics[width=0.49\linewidth]{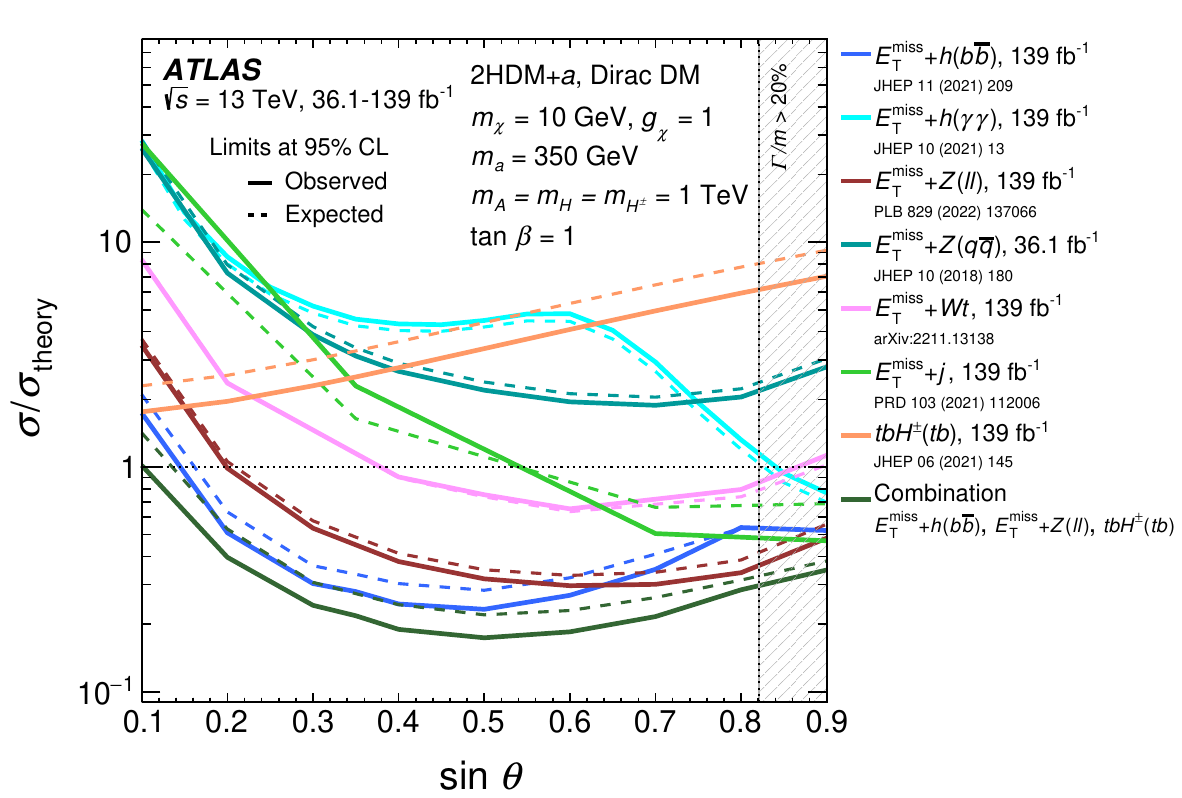}}\\
\subfloat[]{\includegraphics[width=0.49\linewidth]{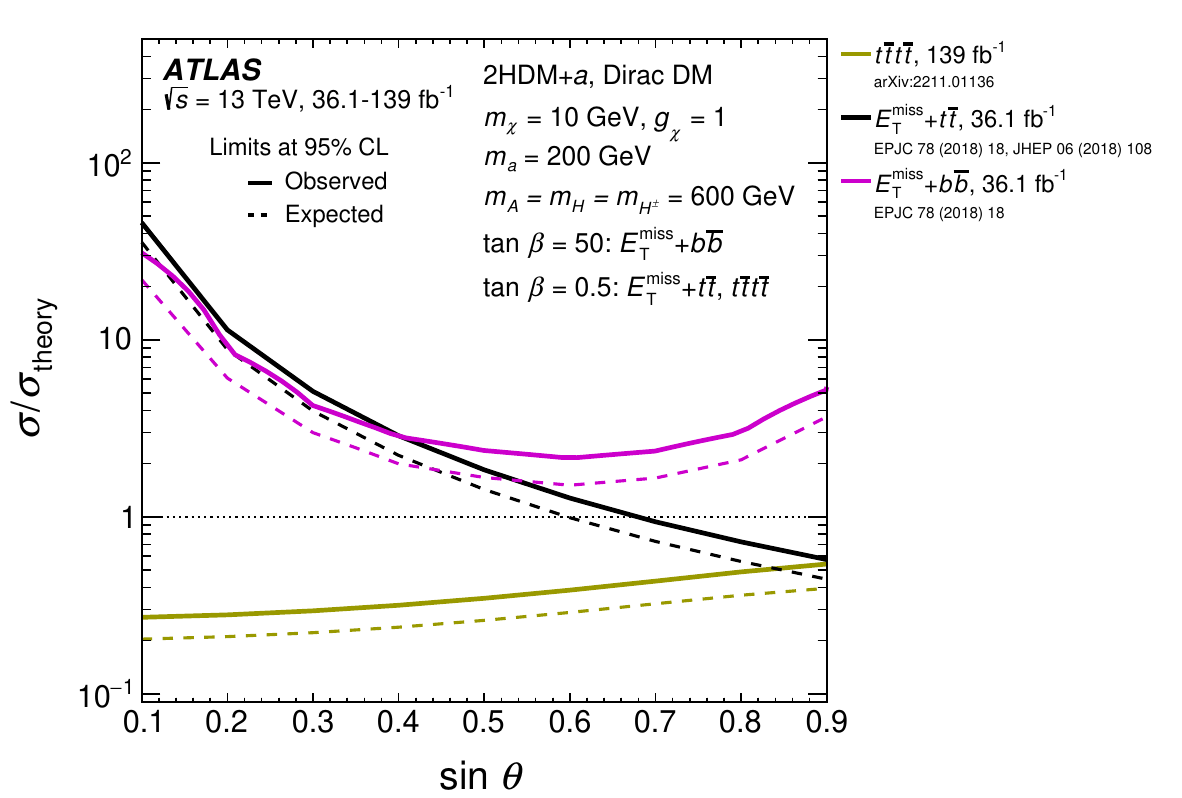}}
\subfloat[]{\includegraphics[width=0.49\linewidth]{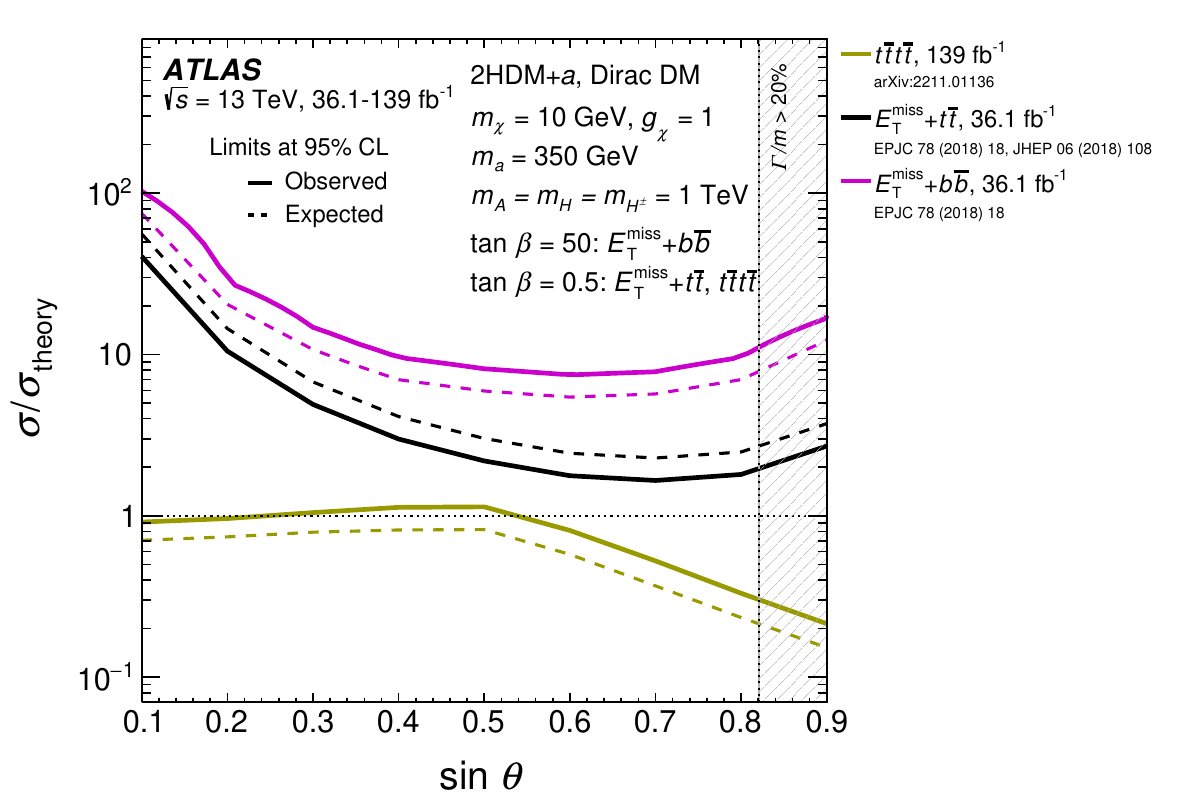}}
\caption{
Observed (solid lines) and expected (dashed lines) exclusion limits at 95\% CL for the \hdma\ as a function of $\sin\theta$.
Results in the subfigures (a) and (b) are derived for the default value $\tan\beta=1$ of Scenario 4,
while those in subfigures (c) and (d) are for alternative values of $\tan\beta=0.5$ or $\tan\beta=50$.
Subfigures (a) and (c) correspond to $m_A = 0.6$~\TeV, $m_a = 200\;\GeV$ (low-mass hypothesis),
while (b) and (d) contain results for $m_A = 1.0$~\TeV, $m_a = 350\;\GeV$ (high-mass hypothesis).
The results are shown for several individual searches and the combination of the \monohbb, \monozll, and \htb searches.
The dashed grey regions indicate the region where the width of any of the Higgs bosons exceeds $20\%$ of its mass.
}
\label{fig:scan_sint}
\end{figure}

\FloatBarrier

\subsection{Scenario 5: variation of \mDM}
\label{sec:thdma_sum_mDM}

In Figure~\ref{fig:scan_mDM}, the sensitivity of the different searches is compared as a function of the DM mass $m_{\chi}$,
which is the parameter with the strongest impact on the relic density predicted by the 2HDM+$a$.
This corresponds to benchmark Scenario 5 in Section~\ref{sec:thdma}.
The sensitivity of the searches is quantified as the observed exclusion limits on the ratio of the excluded cross-section to the nominal cross-section of the model (left vertical axis).
The predicted relic density (right vertical axis) for each value of \mDM\ is overlaid on the plot as a long-dashed line.
The region at $m_\chi = \ma/2 = 200$~\GeV corresponds to the $\pscal$-funnel region~\cite{Djouadi:2005dz,Bagnaschi:2015eha,2HDMWGproxi}
where the predicted relic density is depleted by the resonant enhancement of the process $\chi\bar\chi \rightarrow \pscal \rightarrow \mathrm{SM}$.
A second funnel region at $m_{\chi} = m_{A}/2 = 500$~\GeV, corresponding to the resonant enhancement of the process $\chi\bar\chi \rightarrow A \rightarrow \mathrm{SM}$,
is not fully included in the probed \mDM range but partially visible as a decrease in the predicted relic density for $\mDM > 400$~\GeV.
The plateau for $m_{\chi} > 200$~\GeV is determined by the increase in annihilation cross-section of the DM particles close to threshold for
the processes $\chichi \rightarrow \ttbar$ (if $m_{\chi}  > m_t$) and $\chichi \rightarrow ah$ (if $m_{\chi} > (m_a + m_h)/2$).
For all signatures shown here, the sensitivity is independent of \mDM as long as the pseudo-scalar mediator, whose mass is fixed at 400~\GeV in this benchmark scenario,
is allowed to decay into a $\chi\bar{\chi}$ pair.
The strongest constraints on this region ($m_{\chi} < 200$~\GeV) from individual searches are provided by the \monozll search, which, together with the \monohbb search, excludes this parameter space.
For higher DM masses, the sensitivity of the $\met+X$ searches decreases rapidly,
while that of the \htb and \tttt searches remains nearly constant.
This is because the corresponding signal processes at LO do not involve the DM particle $\chi$, making the signal cross-sections independent of $m_{\chi}$.
For $m_{\chi}>m_a/2$, the strongest constraints are obtained from the \htb search, which probes cross-sections as low as $\sigma/\sigma_{\mathrm{theory}}\approx 2-3$.
Hence none of the searches excludes the 2HDM+$a$ in this mass region for the chosen benchmark scenario.
It is possible to match the observed relic density for $m_{\chi}\approx 170$~\GeV
without changing the collider phenomenology, although this value is disfavoured by the searches in this benchmark scenario.
It should be noted that the relic density considerations serve as a useful means for putting the \thdma predictions in the context of cosmological observations
but should not be understood as strict constraints on the model parameters. This is because the parameter values giving the correct relic density can change either
if the model is modified to include additional physics at higher energy scales or if a different cosmological history is assumed.

\begin{figure}[htbp!]
\centering
\includegraphics[width=0.7\linewidth]{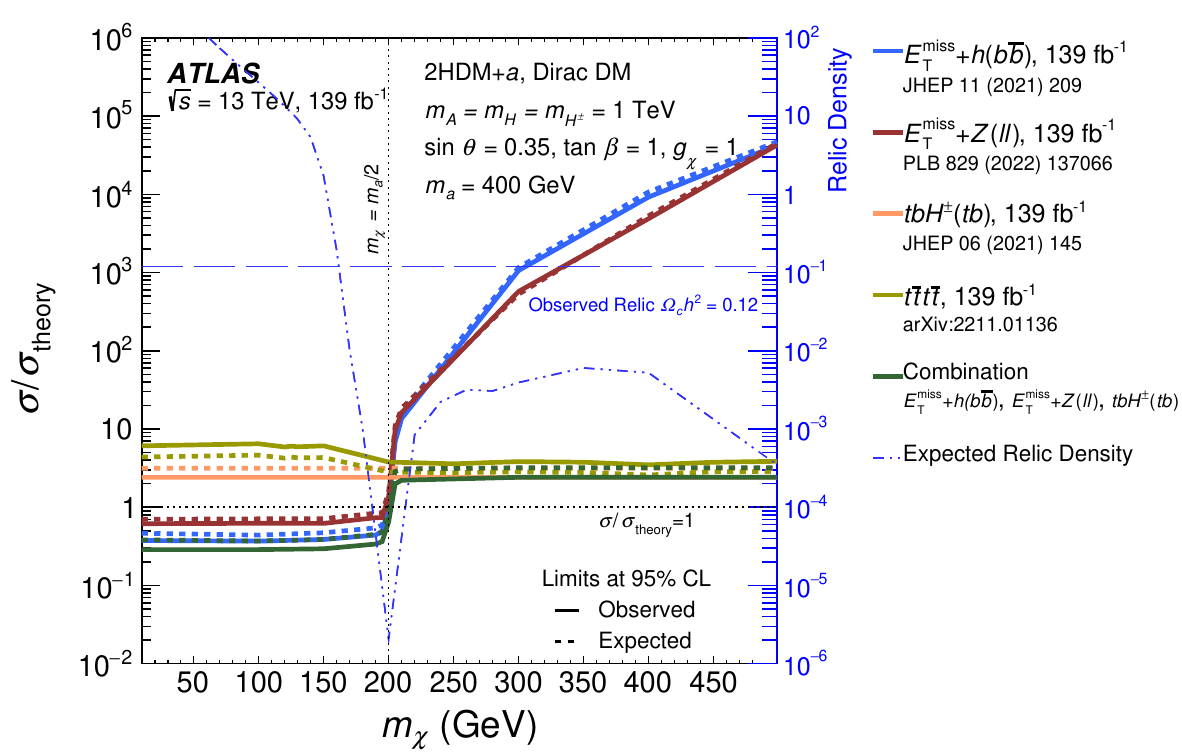}
\caption{
Observed (solid lines) and expected (dashed lines) exclusion limits for the \hdma\ as a function of $m_{\chi}$,
following the parameter choices of $m_{A}=1.0$~\TeV, $m_{a}=400$~\GeV, $\tanb=1.0$, and $\sin\theta=0.35$ (Scenario 5).
The limits are calculated at 95\% CL and are expressed in terms of the ratio of the excluded cross-section to the nominal cross-section of the model.
The results are shown for several individual searches and the combination of the \monozll, \monohbb, and \htb searches.
The relic density for each \mDM\ assumption, calculated with \textsc{MadDM}~\cite{Ambrogi:2018jqj},
is superimposed in the plot (dashed line) and described by the right vertical axis.
The valley at $m_\chi = 200\;\GeV$ indicates the $\pscal$-funnel region~\cite{Djouadi:2005dz,Bagnaschi:2015eha,2HDMWGproxi}
where the predicted relic density is depleted by the resonant enhancement of the processes $\chi\bar\chi \rightarrow A/\pscal \rightarrow \mathrm{SM}$.
}
\label{fig:scan_mDM}
\end{figure}

\FloatBarrier

\subsection{Scenario 6: $\ma-\mDM$ plane}

Exclusion limits as a function of \ma and \mDM corresponding to Scenario 6 are shown in Figure~\ref{fig:scan_ma_mDM}.
The \hlrs searches target the region characterised by $m_a < m_h / 2$, where the decay $h\rightarrow aa$ is kinematically allowed,
and $m_a < 2  m_{\chi}$ where invisible mediator decays are kinematically forbidden.
This region is almost fully excluded by the \hlrs searches under consideration, except for two bands where \ma is close to the masses of the J/$\psi$ and $\Upsilon$ mesons.
As discussed in Section~\ref{subsubsec:hlrs}, these mass regions are excluded from the searches.
Experimentally di-muon searches near the J/$\psi$ mass are challenging, as are $h \rightarrow aa \rightarrow 4g$ searches.
The $\mu^+\mu^-\tau^+\tau^-$ final states~\cite{HIGG-2014-02} have some sensitivity, but are unable to exclude the higher mass region around $m_a=10$~\GeV.
Searches for hadronic final states are complicated by the collimation of the quark pairs,
and dedicated techniques are often required to make signatures such as $b\bar{b}\tau^+\tau^-$ and $b\bar{b}b\bar{b}$ sensitive.
The \hlrs searches are not sensitive for $m_a > m_A / 2$, where invisible mediator decays dominate the branching ratio.
For $m_a < m_h / 2$, this region is excluded by the \hinv\ search.
For larger values of $m_a$, the region $m_a > m_{\chi} / 2$ is excluded by the \monohbb search up to $m_a \approx 600$~\GeV.
The unexcluded high-$m_a$, high-$m_\chi$ region can be probed by searches for the mediator or heavy Higgs boson states in signatures such as \fourtop~and \htb,
which are currently unable to exclude $m_A=1200$~\GeV.

The relic density contour for the case $\Omega_c \mathrm{h}^2 = 0.12$ is superimposed in the plot (long-dashed line).
Regions above this line at low \mDM and below this line at high $m_\chi$, excluding the ``island'' region around ($m_{\chi}\approx 100$~GeV, $m_a\approx 150$~GeV),
have a predicted relic density $\Omega h^2 < 0.12$.
Due to the large Yukawa couping, the annihilation process $\chi\bar\chi \rightarrow \ttbar$ is very efficient.
For regions with light DM (\mchi$< m_t$), this is kinematically inaccessible and the predicted relic density is often over-abundant unless alternative annihilation channels are available.
The most important of these are resonant annihilation when $m_\chi \simeq m_a /2$, and specific decay channels such as $\chi\bar\chi \rightarrow aa$ or $\chi\bar\chi \rightarrow ah$ when allowed or enhanced by kinematics.
For low $m_a$, annihilation to fermions (e.g. \bbbar, \ccbar, $\tau^+\tau^-$) can be efficient enough to overcome their smaller couplings and deplete the relic abundance.
Larger values of \mDM can also satisfy the observed relic density, as these annihilations become more suppressed.

\begin{figure}[htbp!]
\centering
\includegraphics[width=0.7\linewidth]{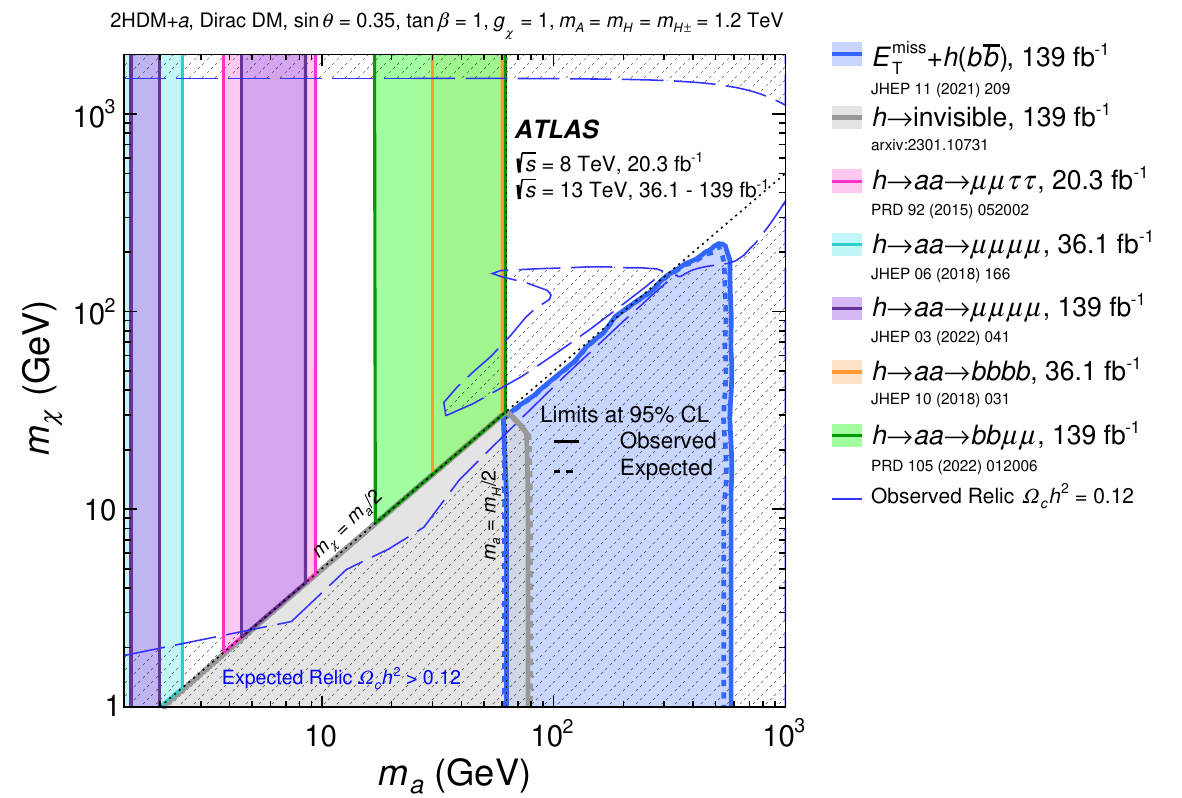}
\caption{
Observed (solid lines) and expected (dashed lines) exclusion regions at 95\% CL in the $(\ma, \mDM)$ plane
following the parameter choices of $m_{A}=1.2$~\TeV, $\tanb=1.0$, and $\sin\theta=0.35$ (Scenario 6).
The relic density contour for the case $\Omega_c \mathrm{h}^2 = 0.12$, calculated with \textsc{MadDM}~\cite{Ambrogi:2018jqj}, is superimposed in the plot (long-dashed line).
The shaded regions mark parameter values for which the model predicts a relic density greater than the observed value $\Omega_c \mathrm{h}^2 = 0.12$.
The ``island'' around ($\mDM\approx 100$~\GeV, $\ma\approx 100$~\GeV) corresponds to the resonant enhancement of
the process $\chi\bar\chi \rightarrow ah \rightarrow \mathrm{SM}$, which depletes the relic density.
}
\label{fig:scan_ma_mDM}
\end{figure}

\FloatBarrier

% End of text imported from the .//sections/results.tex input file

\FloatBarrier

\section{Conclusion}
\label{sec:conclusion}
%

% The next lines are included from the .//sections/conclusion.tex input file
A broad variety of searches for new phenomena performed by the ATLAS Collaboration are summarised
and interpreted in the context of a common LHC dark matter benchmark model,
namely a Two-Higgs-Doublet-Model with an additional pseudo-scalar mediator $a$ (2HDM+$a$),
which couples the dark matter particles to the Standard Model.
This model predicts a rich phenomenology of processes resulting in a diverse range of final-state signatures.
The searches presented provide sensitivity across a wide range of the model parameter space.

The results are based on up to 139~fb$^{-1}$ of proton-proton collision data at a centre-of-mass energy of $\sqrt{s}=13$~\TeV
collected by the ATLAS detector at the LHC in the years 2015--2018, and are in agreement with the Standard Model predictions.
Therefore, the results are translated into exclusion limits on the \thdma for a wide selection of representative benchmark scenarios.
These include previously explored benchmark scenarios based on the recommendations of the LHC Dark Matter Working Group
as well as several new benchmark scenarios that provide further insights into the rich collider phenomenology of the 2HDM+$a$.
All benchmark scenarios rely on the simplifying assumption that the additional Higgs bosons of the 2HDM are mass degenerate ($m_A=m_H=m_{H^{\pm}}$).
The exploration of additional benchmark scenarios in which this assumption is relaxed is left to future publications.

Masses of the pseudo-scalar mediator $a$ are excluded up to 560~\GeV for $m_A=m_H=m_{H^{\pm}}=1.2$~\TeV, $\sin\theta=0.35$, and $\tan\beta=1.0$.
Values of \ma up to 640~GeV are excluded for $m_A=m_H=m_{H^{\pm}}=2.0$~\TeV, $\sint=0.7$, and $\tanb=1.0$.
The \monozll\ and \monohbb\ searches are the most sensitive analyses in this region of large heavy Higgs boson masses ($m_A=m_H=m_{H^{\pm}}$).
These results mark a significant improvement compared to previous results based on only 36~fb$^{-1}$ of proton-proton collision data at the same centre-of-mass energy,
for which the maximum exclusion reach in $m_a$ was up to 340~\GeV for $m_A=m_H=m_{H^{\pm}}=1.0$~\TeV, $\sin\theta=0.35$, and $\tan\beta=1.0$.
These improvements are not only due to the larger amount of data used in this publication but various improvements in the analysis strategies of the individual searches
as well as the statistical combination of the most sensitive individual results.

Additionally, the interpretation of the \htb search in the context of the \thdma allows for values of \mA up to 650~\GeV to be excluded across the full probed \ma range.
The \htb search was not considered in any previous publications but is the most sensitive analysis in this low-\mA region for mediator masses above 400~\GeV.
It also signficantly extends the exclusion reach in \tanb across the full probed \ma range from $\tanb\approx 0.6$ achieved via a \tttt search based on 36~\ifb
to $\tanb\approx 1.2$ for the same benchmark scenario 3a in this publication.
This highlights the importance of searches that are not classically interpreted in the context of DM in constraining more complex models of DM, such as the 2HDM+$a$.
A statistical combination of the \monozll, \monohbb, and \htb searches is performed.
This combination extends the sensitivity to the \thdma compared with the sensitivities derived from the individual searches across different regions of the \thdma parameter space.
Compared to the previous summary of 36~fb$^{-1}$ results, the excluded regions in these benchmark scenarios are considerably extended. Nonetheless, there remains sizeable unexcluded model space, especially if the full parameter space or more general 2HDM+$a$ models are considered.
Finally, for the first time the results of searches targeting \hlrs are used to constrain a part of previously unprobed \thdma parameter space.
The results in this paper represent the most comprehensive set of constraints on the 2HDM+$a$ obtained by the ATLAS Collaboration to date.

% End of text imported from the .//sections/conclusion.tex input file

%
\section*{Acknowledgements}
%

% The next lines are included from the .//acknowledgements/Acknowledgements.tex input file
%
%

%
%

We thank CERN for the very successful operation of the LHC and its injectors, as well as the support staff at
CERN and at our institutions worldwide without whom ATLAS could not be operated efficiently.

The crucial computing support from all WLCG partners is acknowledged gratefully, in particular from CERN, the ATLAS Tier-1 facilities at TRIUMF/SFU (Canada), NDGF (Denmark, Norway, Sweden), CC-IN2P3 (France), KIT/GridKA (Germany), INFN-CNAF (Italy), NL-T1 (Netherlands), PIC (Spain), RAL (UK) and BNL (USA), the Tier-2 facilities worldwide and large non-WLCG resource providers. Major contributors of computing resources are listed in Ref.~\cite{ATL-SOFT-PUB-2023-001}.

We gratefully acknowledge the support of ANPCyT, Argentina; YerPhI, Armenia; ARC, Australia; BMWFW and FWF, Austria; ANAS, Azerbaijan; CNPq and FAPESP, Brazil; NSERC, NRC and CFI, Canada; CERN; ANID, Chile; CAS, MOST and NSFC, China; Minciencias, Colombia; MEYS CR, Czech Republic; DNRF and DNSRC, Denmark; IN2P3-CNRS and CEA-DRF/IRFU, France; SRNSFG, Georgia; BMBF, HGF and MPG, Germany; GSRI, Greece; RGC and Hong Kong SAR, China; ISF and Benoziyo Center, Israel; INFN, Italy; MEXT and JSPS, Japan; CNRST, Morocco; NWO, Netherlands; RCN, Norway; MEiN, Poland; FCT, Portugal; MNE/IFA, Romania; MESTD, Serbia; MSSR, Slovakia; ARRS and MIZ\v{S}, Slovenia; DSI/NRF, South Africa; MICINN, Spain; SRC and Wallenberg Foundation, Sweden; SERI, SNSF and Cantons of Bern and Geneva, Switzerland; MOST, Taipei; TENMAK, T\"urkiye; STFC, United Kingdom; DOE and NSF, United States of America.

Individual groups and members have received support from BCKDF, CANARIE, CRC and DRAC, Canada; CERN-CZ, PRIMUS 21/SCI/017 and UNCE SCI/013, Czech Republic; COST, ERC, ERDF, Horizon 2020, ICSC-NextGenerationEU and Marie Sk{\l}odowska-Curie Actions, European Union; Investissements d'Avenir Labex, Investissements d'Avenir Idex and ANR, France; DFG and AvH Foundation, Germany; Herakleitos, Thales and Aristeia programmes co-financed by EU-ESF and the Greek NSRF, Greece; BSF-NSF and MINERVA, Israel; Norwegian Financial Mechanism 2014-2021, Norway; NCN and NAWA, Poland; La Caixa Banking Foundation, CERCA Programme Generalitat de Catalunya and PROMETEO and GenT Programmes Generalitat Valenciana, Spain; G\"{o}ran Gustafssons Stiftelse, Sweden; The Royal Society and Leverhulme Trust, United Kingdom.

In addition, individual members wish to acknowledge support from Chile: Agencia Nacional de Investigaci\'on y Desarrollo (FONDECYT 1190886, FONDECYT 1210400, FONDECYT 1230812, FONDECYT 1230987); China: National Natural Science Foundation of China (NSFC - 12175119, NSFC 12275265, NSFC-12075060); Czech Republic: PRIMUS Research Programme (PRIMUS/21/SCI/017); European Union: European Research Council (ERC - 948254), Horizon 2020 Framework Programme (MUCCA - CHIST-ERA-19-XAI-00), European Union, Future Artificial Intelligence Research (FAIR-NextGenerationEU PE00000013), Italian Center for High Performance Computing, Big Data and Quantum Computing (ICSC, NextGenerationEU), Marie Sklodowska-Curie Actions (EU H2020 MSC IF GRANT NO 101033496); France: Agence Nationale de la Recherche (ANR-20-CE31-0013, ANR-21-CE31-0013, ANR-21-CE31-0022), Investissements d'Avenir Idex (ANR-11-LABX-0012), Investissements d'Avenir Labex (ANR-11-LABX-0012); Germany: Baden-Württemberg Stiftung (BW Stiftung-Postdoc Eliteprogramme), Deutsche Forschungsgemeinschaft (DFG - CR 312/5-1); Italy: Istituto Nazionale di Fisica Nucleare (FELLINI G.A. n. 754496, ICSC, NextGenerationEU); Japan: Japan Society for the Promotion of Science (JSPS KAKENHI JP21H05085, JSPS KAKENHI JP22H01227, JSPS KAKENHI JP22H04944); Netherlands: Netherlands Organisation for Scientific Research (NWO Veni 2020 - VI.Veni.202.179); Norway: Research Council of Norway (RCN-314472); Poland: Polish National Agency for Academic Exchange (PPN/PPO/2020/1/00002/U/00001), Polish National Science Centre (NCN 2021/42/E/ST2/00350, NCN UMO-2019/34/E/ST2/00393, UMO-2020/37/B/ST2/01043, UMO-2021/40/C/ST2/00187); Slovenia: Slovenian Research Agency (ARIS grant J1-3010); Spain: BBVA Foundation (LEO22-1-603), Generalitat Valenciana (Artemisa, FEDER, IDIFEDER/2018/048), La Caixa Banking Foundation (LCF/BQ/PI20/11760025), Ministry of Science and Innovation (MCIN \& NextGenEU -PCI2022-135018-2, MICIN \& FEDER -PID2021-125273NB, RYC2019-028510-I, RYC2020-030254-I, RYC2021-031273-I, RYC2022-038164-I), PROMETEO and GenT Programmes Generalitat Valenciana (CIDEGENT/2019/023, CIDEGENT/2019/027); Sweden: Swedish Research Council (VR 2018-00482, VR 2022-03845, VR 2022-04683, VR grant 2021-03651), Knut and Alice Wallenberg Foundation (KAW 2017.0100, KAW 2018.0157, KAW 2018.0458, KAW 2019.0447); Switzerland: Swiss National Science Foundation (SNSF - PCEFP2\_194658); United Kingdom: Leverhulme Trust (Leverhulme Trust RPG-2020-004); United States of America: Neubauer Family Foundation.

%
%

% End of text imported from the .//acknowledgements/Acknowledgements.tex input file

%
%
%
%
%
%

%
%

%
%
%
\printbibliography
\clearpage
% ATLAS Collaboration author list
% Reference date of EXOT-2018-64 is 2023-04-06
% Author list last updated on date 15-MAY-24
% Data extracted on 15-May-2024 for paper reference EXOT-2018-64
% at 10:03pm
 
\begin{flushleft}
\hypersetup{urlcolor=black}
{\Large The ATLAS Collaboration}

\bigskip

\AtlasOrcid[0000-0002-6665-4934]{G.~Aad}$^\textrm{\scriptsize 102}$,
\AtlasOrcid[0000-0002-5888-2734]{B.~Abbott}$^\textrm{\scriptsize 120}$,
\AtlasOrcid[0000-0002-1002-1652]{K.~Abeling}$^\textrm{\scriptsize 55}$,
\AtlasOrcid[0000-0001-5763-2760]{N.J.~Abicht}$^\textrm{\scriptsize 49}$,
\AtlasOrcid[0000-0002-8496-9294]{S.H.~Abidi}$^\textrm{\scriptsize 29}$,
\AtlasOrcid[0000-0002-9987-2292]{A.~Aboulhorma}$^\textrm{\scriptsize 35e}$,
\AtlasOrcid[0000-0001-5329-6640]{H.~Abramowicz}$^\textrm{\scriptsize 151}$,
\AtlasOrcid[0000-0002-1599-2896]{H.~Abreu}$^\textrm{\scriptsize 150}$,
\AtlasOrcid[0000-0003-0403-3697]{Y.~Abulaiti}$^\textrm{\scriptsize 117}$,
\AtlasOrcid[0000-0003-0762-7204]{A.C.~Abusleme~Hoffman}$^\textrm{\scriptsize 137a}$
% ...
\bigskip
\\

% ...
$^{29}$Physics Department, Brookhaven National Laboratory, Upton NY; United States of America.\\
% ...
$^{35}$$^{(a)}$Facult\'e des Sciences Ain Chock, R\'eseau Universitaire de Physique des Hautes Energies - Universit\'e Hassan II, Casablanca;$^{(b)}$Facult\'{e} des Sciences, Universit\'{e} Ibn-Tofail, K\'{e}nitra;$^{(c)}$Facult\'e des Sciences Semlalia, Universit\'e Cadi Ayyad, LPHEA-Marrakech;$^{(d)}$LPMR, Facult\'e des Sciences, Universit\'e Mohamed Premier, Oujda;$^{(e)}$Facult\'e des sciences, Universit\'e Mohammed V, Rabat;$^{(f)}$Institute of Applied Physics, Mohammed VI Polytechnic University, Ben Guerir; Morocco.\\
% ...
$^{49}$Fakult\"{a}t Physik , Technische Universit{\"a}t Dortmund, Dortmund; Germany.\\
% ...
$^{55}$II. Physikalisches Institut, Georg-August-Universit\"{a}t G\"ottingen, G\"ottingen; Germany.\\
% ...
$^{102}$CPPM, Aix-Marseille Universit\'e, CNRS/IN2P3, Marseille; France.\\
% ...
$^{117}$Department of Physics, New York University, New York NY; United States of America.\\
% ...
$^{120}$Homer L. Dodge Department of Physics and Astronomy, University of Oklahoma, Norman OK; United States of America.\\
% ...
$^{137}$$^{(a)}$Departamento de F\'isica, Pontificia Universidad Cat\'olica de Chile, Santiago;$^{(b)}$Millennium Institute for Subatomic physics at high energy frontier (SAPHIR), Santiago;$^{(c)}$Instituto de Investigaci\'on Multidisciplinario en Ciencia y Tecnolog\'ia, y Departamento de F\'isica, Universidad de La Serena;$^{(d)}$Universidad Andres Bello, Department of Physics, Santiago;$^{(e)}$Instituto de Alta Investigaci\'on, Universidad de Tarapac\'a, Arica;$^{(f)}$Departamento de F\'isica, Universidad T\'ecnica Federico Santa Mar\'ia, Valpara\'iso; Chile.\\
% ...
$^{150}$Department of Physics, Technion, Israel Institute of Technology, Haifa; Israel.\\
$^{151}$Raymond and Beverly Sackler School of Physics and Astronomy, Tel Aviv University, Tel Aviv; Israel.\\
% ...
$^{a}$ Also Affiliated with an institute covered by a cooperation agreement with CERN.\\
$^{b}$ Also at An-Najah National University, Nablus; Palestine.\\
$^{c}$ Also at Borough of Manhattan Community College, City University of New York, New York NY; United States of America.\\
$^{d}$ Also at Center for High Energy Physics, Peking University; China.\\
$^{e}$ Also at Center for Interdisciplinary Research and Innovation (CIRI-AUTH), Thessaloniki; Greece.\\
$^{f}$ Also at Centro Studi e Ricerche Enrico Fermi; Italy.\\
% ...

\end{flushleft}

% Created with Glance <Atlas.Glance@cern.ch>

%

%
%

%

%
%
%
%

\end{document}